\documentclass[%
 amsmath,amssymb,
 aps,
 prd,
twocolumn, superscriptaddress
]{revtex4-1}

\usepackage[normalem]{ulem}

\usepackage{graphicx}% Include figure files
\usepackage{dcolumn}% Align table columns on decimal point
\usepackage{bm}% bold math
\usepackage{hyperref}% add hypertext capabilities
\usepackage{xcolor}

\usepackage{color}
\definecolor{DarkGreen}{rgb}{.1, .5, .1}

\begin{document}

\preprint{APS/123-QED}

\title{Equilibrium, radial stability and non-adiabatic gravitational collapse of anisotropic neutron stars}% Force line breaks with \\

%\author{Juan Z\'arate P.}
\author{Juan M. Z. Pretel}
 \email{juanzarate@if.ufrj.br}%Lines break automatically or can be forced with \\
 \affiliation{
 Instituto de F\'\i sica, Universidade Federal do Rio de Janeiro,\\
 CEP 21941-972 Rio de Janeiro, RJ, Brazil 
}

\date{\today}% It is always \today, today,

\begin{abstract}
In this work we construct families of anisotropic neutron stars for an equation of state compatible with the constraints of the gravitational-wave event GW170817 and for four anisotropy ansatze. Such stars are subjected to a radial perturbation in order to study their stability against radial oscillations and we develop a dynamical model to describe the non-adiabatic gravitational collapse of the unstable anisotropic configurations whose ultimate fate is the formation of a black hole. We find that the standard criterion for radial stability $dM/d\rho_c >0$ is not always compatible with the calculation of the oscillation frequencies for some anisotropy ansatze, and each anisotropy parameter is constrained taking into account the recent restriction of maximum mass of neutron stars. We further generalize the TOV equations within a non-adiabatic context and we investigate the dynamical behaviour of the equation of state, heat flux, anisotropy factor and mass function as an unstable anisotropic star collapses. After obtaining the evolution equations we recover, as a static limit, the background equations.
\end{abstract}

\maketitle

%\tableofcontents

\section{Introduction} 

The most common matter-energy distribution for modeling the internal structure of compact stars is an isotropic perfect fluid. Nevertheless, there are strong arguments suggesting that nuclear matter at very high densities and pressures could naturally be described by an anisotropic fluid, that is, when the radial and tangential components of the pressure are not equal. As a matter of fact, the anisotropy could be generated by the presence of strong magnetic fields \cite{Chaichian, Ferrer2010, Yazadjiev2012, Folomeev2015}, solid cores \cite{Ruderman1972, Kippenhahn}, superfluidity \cite{Carter1998, Heiselberg2000}, pion condensed phase configurations in neutron stars \cite{Sawyer1972, Sawyer1973}, etc. Furthermore, it is possible to obtain an anisotropic perfect fluid by combining the energy-momentum tensors of two isotropic perfect fluids \cite{Letelier1980, HerreraSantos1997, Rezzolla}.

As widely reported in the literature \cite{BowersLiang, Heintzmann1975, Gleiser2002, Horvat2011, HerreraBarreto, Maurya2015, ArbanilMalheiro, Ivanov2017, Maurya2018, Delgado2018, Maurya2019, Ortiz2019}, the presence of anisotropy affects a number of important physical properties of compact stars such as the mass-radius relation, compactness, surface redshift, moment of inertia as well as the scalarization in scalar-tensor theories \cite{Silva2015}. Indeed, the equation of state (EoS) plays a fundamental role in determining the internal structure of such stars and, consequently, in imposing stability limits. Therefore, it is important to carry out a stability analysis of anisotropic neutron stars taking into account the LIGO-Virgo constraints on the EoS for nuclear matter as a result of observation of the event GW170817 --- the first direct detection of gravitational waves from the coalescence of a neutron star binary system \cite{Abbott1, Abbott2}. In that respect, we are interested in considering a realistic EoS, which is compatible with the restriction obtained from this merger, and exploring the effects it can have on the physical characteristics of stable and unstable anisotropic stars.

It is a well known fact that the solutions of the Tolman-Oppenheimer-Volkov (TOV) equations describe stellar configurations in hydrostatic equilibrium. Nonetheless, such equilibrium can be stable or unstable with respect to a compression or decompression caused by radial perturbations. Indeed, a conventional technique widely used to indicate the onset of instability is the $M(\rho_c)$ method \cite{Horvat2011}, also known as a necessary condition for stability analysis of compact stars \cite{Glendenning, Haensel2007}. This boundary between the stable and unstable stars describes the maximum amount of mass that can exist in a configuration before it must undergo a gravitational collapse. On the other hand, a sufficient condition for stability is to calculate the frequencies of the normal radial modes of relativistic vibrations \cite{Glendenning, Haensel2007, Bardeen1966}, where Einstein field equations have to be linearized around the equilibrium configuration. If any of these squared frequencies is positive we have stable radial oscillations, whereas negative squared frequencies imply increasing or decreasing perturbations with time, i.e. the star is unstable.

In general relativity (GR), the stability analysis for isotropic compact stars with respect to radial perturbations has been widely discussed in the literature \cite{Chandrasekhar1, Chandrasekhar2, Chanmugam1977, Benvenuto1991, VeC1992, Gondek, KokkotasR2001, FloresL2010, PanotopoulosL2017, PanotopoulosL2018, Pereira2018, Sagun2020, Clemente2020, Pretel2020, Arbanil2020}, whereas within the context of anisotropic configurations the normal radial modes technique has been used only in some specific cases, see e.g., \cite{Hillebrandt1976, Gleiser2003, Karlovini2004, Horvat2011, ArbanilMalheiro, Isayev2017}. In particular, for anisotropic strange stars described by the MIT bag model EoS, it was shown that the $M(\rho_c)$ method is not compatible with the calculation of frequencies to predict the onset of instability \cite{ArbanilMalheiro} for the anisotropy profile proposed by Bowers and Liang \cite{BowersLiang}, this is, the maximum-mass stellar configurations do not correspond to the zero squared frequencies of the fundamental mode. Therefore, it is required to calculate the frequency of the oscillation modes in order to have absolute certainty about the radial stability of an anisotropic neutron star. But what could be the origin of the stellar oscillations? Just as the oscillations inside the Earth are excited by earthquakes and used in seismology to study the structure of the Earth, the fluid pulsations in neutron stars can be excited by cracks in the crust (a starquake) \cite{Franco2000}, as well as by tidal effects in binary inspirals \cite{Hinderer2016, Chirenti2017}.

In the present paper, in addition to generating hydrostatically stable anisotropic configurations, we are also interested in studying the process of gravitational collapse of the unstable configurations. In this regard, the pioneering work about gravitational collapse for a spherically symmetric distribution of matter in the form of dust cloud was carried out by Oppenheimer and Snyder \cite{Oppenheimer1939}. Such idealized treatment has been improved by introducing a pressure gradient \cite{Misner1964} and by replacing the exterior Schwarzschild metric by the Vaidya one \cite{Vaidya}. In this way, the relativistic equations for the adiabatic, spherically symmetric gravitational collapse as given in Ref. \cite{Misner1964} were modified by Misner \cite{Misner1965} in order to allow an extremely simplified heat-transfer process (where the internal energy is converted into an outward flux of neutrinos). Moreover, the equations that govern the gravitational collapse of a ball of charged perfect fluid were derived by Bekenstein \cite{Bekenstein1971}.

Over the years, it has been possible to construct physically viable gravitational collapse scenarios for isotropic fluids that include dissipative fluxes such as heat flow \cite{Santos1985, Oliveira1985, Herrera1989, Bonnor1989, Herrera2006, Ivanov2012} and shear and bulk viscosities \cite{Chan1994}. In turn, gravitational collapse models have also been developed for anisotropic fluids with dissipative processes \cite{Martinez1996, Chan2001, Reddy2015, Veneroni2019, Pretel2019, Govender2019}, in the presence of electromagnetic field  \cite{Prisco2007, Pinheiro2013, Ivanov2019a, Ivanov2019b, Bhatti2020} and with cosmological constant \cite{Mahomed2020a, Mahomed2020b}. Particularly the gravitational collapse of neutron stars considered as initial configurations has been investigated by the authors in Refs. \cite{Martinez1996, Pretel2019, Oliveira1986, Martinez1994}, however, they did not perform an a priori analysis on the stellar stability of the initial static Schwarzschild configurations against radial pulsations. A complete analysis on stability and gravitational collapse for isotropic fluids was conducted by Ghezzi \cite{Ghezzi2005}  (where the charged neutron stars in the unstable branch collapse directly to form black holes), and more recently in Ref. \cite{Pretel2020} for neutron stars with realistic EoSs.

We construct families of anisotropic neutron stars based on an EoS compatible with the recent observations, and we obtain the oscillation spectrum for each family by means of radial perturbations. In addition, we investigate the dynamical evolution of unstable anisotropic stars whose final fate is the formation of a black hole as a consequence of a non-adiabatic gravitational collapse. The spherical surface of the collapsing star divides spacetime into two different four-dimensional manifolds; an interior region with heat flux --- described by a shear-free line element --- and an exterior region which is described by the Vaidya metric for pure outgoing radiation. The paper is organized as follows: In Sec. \ref{Sec2} we present the basic formalism to describe equilibrium configurations for four anisotropy ansatze and deal with the normal radial modes. In Sec. \ref{Sec3} we develop a gravitational collapse model by introducing a time dependency on the metric functions that allows us to recover the static case under a certain limit. In Sec. \ref{Sec4} we present a discussion of the numerical results. The paper ends with our conclusions in Sec. \ref{Sec5}. We adopt the signature $(-, +, +, +)$ and physical units will be used throughout this work.

%--------------------------------------------------------------

\section{Basic equations for stellar structure}\label{Sec2}

The line element describing the interior spacetime of a spherically symmetric star, is written in the well-known form
\begin{equation}\label{1}
ds^2 = g_{\mu\nu}dx^\mu dx^\nu = -e^{2\psi}(dx^0)^2 + e^{2\lambda} dr^2 + r^2d\Omega^2,  \ \
\end{equation}

\noindent
where $x^\mu = (ct, r, \theta, \phi)$, and $d\Omega^2 = d\theta^2 + \sin^2\theta d\phi^2$ is the line element on the unit 2-sphere. The metric functions $\psi$ and $\lambda$, in principle, depend on $x^0$ and $r$.

With regard to the matter-energy distribution, we assume that the system is composed of an anisotropic perfect fluid, where the components of the pressure are not equal to each other, namely \cite{BowersLiang, HerreraBarreto, DonevaYazadjiev}
\begin{equation}\label{2}
T_{\mu\nu} = (\epsilon + p_t) u_\mu u_\nu + p_t g_{\mu\nu} - \sigma k_\mu k_\nu ,
\end{equation}
with $u^\mu$ being the four-velocity of the fluid, $\epsilon = c^2\rho$ the energy density (where $\rho$ denotes mass density), $\sigma \equiv p_t - p_r$ the anisotropy factor, $p_r$ the radial pressure, $p_t$ the tangential pressure, and $k^\mu$ is a unit spacelike four-vector. The four-vectors $u^\mu$ and $k^\mu$ must satisfy the following properties
\begin{equation}\label{3}
u_\mu u^\mu = -1,   \qquad   k_\mu k^\mu = 1,   \qquad   u_\mu k^\mu = 0.
\end{equation}

Within the context of GR, the spacetime geometry and the matter-energy distribution are related by the Einstein field equations
\begin{equation}\label{4}
    G_{\mu\nu} \equiv R_{\mu\nu} - \frac{1}{2}g_{\mu\nu}R = \kappa T_{\mu\nu} ,
\end{equation}
where $\kappa \equiv 8\pi G/c^4$, $G_{\mu\nu}$ is the Einstein tensor, $R_{\mu\nu}$ the Ricci tensor, and $R$ denotes the scalar curvature. Here $G$ is the gravitational constant and $c$ is the speed of light in physical units.

\subsection{Background and TOV equations}

In the case of hydrostatic equilibrium none of the metric and thermodynamic quantities depends on the time coordinate $x^0$, which entails that $k^\mu = (0, e^{-\lambda}, 0, 0)$ and hence the energy-momentum tensor contains only non-zero diagonal components $T_{\mu}^{\ \nu} = {\rm diag}(-\epsilon, p_r, p_t, p_t)$. Consequently, from Eqs. (\ref{1})-(\ref{4}) together with the conservation law of energy and momentum, the relativistic structure of an anisotropic star in the state of hydrostatic equilibrium is governed by the TOV equations
\begin{eqnarray}
\frac{dm}{dr} & = & 4\pi r^2 \rho ,  \label{5}  \\
\frac{dp_r}{dr} & = & -\left[ \frac{p_r + c^2\rho}{c^2} \right]\left[ \frac{Gm}{r^2} + \frac{4\pi G}{c^2}rp_r \right] \left[ 1 - \frac{2Gm}{c^2r} \right]^{-1}  \nonumber  \\
&& + \frac{2}{r}\sigma,  \label{6}  \\
\frac{d\psi}{dr} & = & -\frac{1}{p_r + c^2\rho}\frac{dp_r}{dr} + \frac{2 \sigma}{r(p_r + c^2\rho)} ,  \label{7}
\end{eqnarray}
where $m(r)$ is the mass enclosed in the sphere of radius $r$. The metric function $\lambda(r)$ is obtained by means of the relation
\begin{equation}\label{8}
    e^{-2\lambda} = 1 - \frac{2Gm}{c^2r} .
\end{equation}
As usual, we define the radius of the star when the radial pressure vanishes, i.e., the surface of the anisotropic star is reached when $p_r(r = R) = 0$, and the total gravitational mass of the star is given by $M \equiv m(R)$.

Given a barotropic EoS of the form $p_r = p_r(\rho)$ and a defined anisotropy relation for $\sigma$, Eqs. (\ref{5}) and (\ref{6}) can be integrated for a given central density and by guaranteeing regularity at the center of the star. Besides that, since the equilibrium system is a spherically symmetric star, the exterior spacetime of the anisotropic fluid must be described by the Schwarzschild metric so that the continuity of the metric on the surface imposes another boundary condition for the differential equation (\ref{7}). Thus, the system of Eqs. (\ref{5})-(\ref{7}) is solved under the requirement of the following boundary conditions
\begin{equation}\label{9}
\rho(0) = \rho_c,   \ \quad   m(0) = 0,   \ \quad   \psi(R) = \frac{1}{2}\ln\left[ 1 - \frac{2GM}{c^2R} \right] . \ \
\end{equation}

\subsection{Stability criterion through adiabatic radial oscillations}

In order to study the (in)stability of anisotropic neutron stars, it is necessary to calculate the frequencies of normal vibration modes. In fact, this involves an examination of radial perturbations in anisotropic fluid configurations. For such analysis we consider adiabatic vibrations, that is, we shall neglect the heat transfer between neighboring fluid elements. 

Oscillation frequencies about the equilibrium state can be found by considering small deviations with respect to the state of hydrostatic equilibrium. In other words, the equilibrium configuration governed by the TOV equations (\ref{5})-(\ref{7}) is radially perturbed in such a way that its spherical symmetry is maintained. Such a perturbation will cause motions in the radial directions so that a fluid element located at radial coordinate $r$ in the unperturbed configuration is displaced to radial coordinate $r+ \xi(x^0,r)$ in the perturbed configuration, where $\xi$ is the Lagrangian displacement. This involves solving the perturbed Einstein equations $\delta G_{\mu\nu} = \kappa \delta T_{\mu\nu}$ for small radial oscillations from equilibrium. Therefore, in order to do a tractable analysis of the pulsations, all equations are linearized in the Eulerian perturbation functions $\delta h$, where the quantity $h$ represents any metric or fluid variable and is decomposed as $h(x^0, r) = h_0(r) + \delta h(x^0, r)$. The quantities denoted by a subscript zero stand for the solutions in the equilibrium configuration.

It is important to note here that the relation between Eulerian and Lagrangian perturbations is given as follows
\begin{equation}\label{10}
 \Delta h(x^0, r) \equiv h[x^0, r+ \xi(x^0, r)] - h_0(r) \cong \delta h + \frac{dh_0}{dr}\xi ,  \ \
\end{equation}
where $\Delta h$ is the Lagrangian perturbation, that is, the change measured by an observer who moves with the fluid.

If in the perturbed state we define $v \equiv \partial r/\partial x^0 = \partial\xi/\partial x^0$, to first order in $\xi$ the non-zero components of the energy-momentum tensor (\ref{2}) take the form
\begin{eqnarray}\label{11}
T_{00} &=& e^{2\psi}\epsilon ,   \ \ \quad   T_{0r} = T_{r0} = -(\epsilon_0 + p_{r0})e^{2\lambda_0}v ,  \nonumber  \\
T_{rr} &=& e^{2\lambda}p_r ,   \ \quad   T_{\theta\theta} = r^2p_t  ,   \ \quad   T_{\phi\phi} = r^2p_t\sin^2\theta ,  \qquad 
\end{eqnarray}
where the four-velocity and the unit four-vector are given by $u^\mu = (e^{-\psi}, ve^{-\psi_0}, 0, 0)$ and $k^\mu = (ve^{\lambda-2\psi}, e^{-\lambda}, 0, 0)$, respectively. Afterward, by retaining the terms only of first order in $\xi$ and $\delta h$, the linearized field equations are 
\begin{eqnarray}
\delta\lambda &=& -\xi \frac{d}{dr}(\psi_0 + \lambda_0) = -\frac{\kappa}{2}r(\epsilon_0 + p_{r0})e^{2\lambda_0}\xi , \qquad \  \label{12}  \\
\frac{\partial(\delta \psi)}{\partial r} &=& \left[ \frac{\delta p_r}{\epsilon_0 + p_{r0}} - \left( 2\frac{d\psi_0}{dr} + \frac{1}{r} \right)\xi  \right]  \nonumber  \\
&&\times \frac{d}{dr}(\psi_0 + \lambda_0) , \label{13}   \\
\delta\epsilon &=&  -\frac{1}{r^2}\frac{\partial}{\partial r}\left[ r^2(\epsilon_0 + p_{r0})\xi \right] = -\xi\frac{d\epsilon_0}{dr}  \nonumber  \\
&& - (\epsilon_0 + p_{r0})\frac{e^{\psi_0}}{r^2}\frac{\partial}{\partial r}(r^2\xi e^{-\psi_0}) - \frac{2}{r}\xi\sigma_0 .  \label{14}
\end{eqnarray}

In addition, the $\mu = r$ component for the four-divergence of the energy-momentum tensor $\nabla_\nu T_\mu^{\ \nu} =0$, provides the following relation
\begin{eqnarray}\label{15}
&& \frac{\partial T_r^{\ 0}}{\partial x^0} + \frac{\partial T_r^{\ r}}{\partial r} + T_r^{\ 0}\frac{\partial}{\partial x^0}(\psi + \lambda)  \nonumber  \\
&& \hspace{1.6cm} + (T_r^{\ r} - T_0^{\ 0})\frac{\partial\psi}{\partial r} + \frac{2}{r}(T_r^{\ r} - p_t) = 0 ,   \qquad
\end{eqnarray}
or alternatively,
\begin{eqnarray}\label{16}
&& (\epsilon_0 + p_{r0})e^{2(\lambda_0 - \psi_0)}\frac{\partial v}{\partial x^0} + \frac{\partial(\delta p_r)}{\partial r} + (\epsilon_0 + p_{r0})\frac{\partial(\delta \psi)}{\partial r}  \nonumber  \\
&& \hspace{1.6cm} + (\delta p_r + \delta\epsilon)\frac{d\psi_0}{dr} - \frac{2}{r}\delta\sigma = 0 .
\end{eqnarray}

Through Einstein equations it was possible to express some perturbations in terms of the Lagrangian displacement $\xi$ and the unperturbed variables. Nonetheless, we still need to have an expression for the perturbation $\delta p_r$, and to obtain it, an additional condition is necessary (the conservation of the baryon number). If $n$ is the number of baryons per unit volume, its conservation in GR is given by $\nabla_\mu J^\mu = 0$, where $J^\mu \equiv nu^\mu$ is the baryon number current. At the same time, if we consider that the EoS has the general structure $n = n(\epsilon, p_r)$, we have 
\begin{equation}\label{17}
\delta p_r = -\xi\frac{dp_{r0}}{dr} - \gamma p_{r0}\frac{e^{\psi_0}}{r^2}\frac{\partial}{\partial r}(r^2\xi e^{-\psi_0}) + \frac{2}{r}\sigma_0\xi\frac{\partial p_r}{\partial\epsilon} ,  \ \
\end{equation}
where $\gamma \equiv \frac{n}{p_r}\frac{dp_r}{dn}$ is a dimensionless quantity that determines the changes of radial pressure associated with variations in the particle number density.  If the entropy is conserved, we obtain the adiabatic index given by 
\begin{eqnarray}\label{18}
\gamma &=& \frac{1}{p_r(\partial n/\partial p_r)}\left[ n - (\epsilon + p_r)\frac{\partial n}{\partial\epsilon} \right]  \nonumber  \\
 &=& \left( 1+ \frac{\epsilon}{p_r} \right)\frac{dp_r}{d\epsilon} .
\end{eqnarray}

We now assume that the Lagrangian displacement has a harmonic time dependence as $\xi(x^0, r) = \xi(r) e^{i\alpha x^0}$ where $c\alpha \equiv \omega$ is a characteristic frequency to be determined. The same procedure is applied for the metric functions and thermodynamic quantities. Thus, the substitution of Eq. (\ref{13}) into (\ref{16}), leads to
\begin{eqnarray}\label{19}
&& \alpha^2(\epsilon_0 + p_{r0})e^{2(\lambda_0 - \psi_0)}\xi = \frac{d(\delta p_r)}{dr} + \delta p_r \frac{d}{dr}(2\psi_0 + \lambda_0)   \nonumber   \\
&& \hspace{0.5cm} - (\epsilon_0 + p_{r0})\left( 2\frac{d\psi_0}{dr} + \frac{1}{r} \right)\left( \frac{d\psi_0}{dr} + \frac{d\lambda_0}{dr} \right)\xi  \nonumber  \\
&& \hspace{0.5cm} + \frac{d\psi_0}{dr}\delta\epsilon - \frac{2}{r}\delta\sigma .
\end{eqnarray}
By means of Eqs. (\ref{7}), (\ref{14}), (\ref{17}) and the $\theta\theta$-component of the field equations, the last expression can be written in terms of the unperturbed variables
\begin{eqnarray}\label{20}
&& \alpha^2(\epsilon_0 + p_{r0})e^{2(\lambda_0 - \psi_0)}\xi = \Delta p_r \frac{d}{dr}(2\psi_0 + \lambda_0) + \frac{d}{dr}(\Delta p_r)   \nonumber  \\
&& \hspace{0.8cm} + \kappa p_t e^{2\lambda_0}(\epsilon_0 + p_{r0})\xi - \xi(\epsilon_0 + p_{r0})\left( \frac{d\psi_0}{dr} \right)^2    \nonumber   \\ 
&& \hspace{0.8cm} + \frac{4\xi}{r}\frac{dp_{r0}}{dr} - \frac{2\sigma_0\xi}{r}\left[ \frac{d}{dr}(2\psi_0 + \lambda_0) + \frac{4}{r} \right]  \nonumber  \\
&& \hspace{0.8cm} - \frac{d}{dr}\left[ \frac{2\sigma_0\xi}{r} \right] - \frac{2}{r}\delta\sigma .
\end{eqnarray} 

Because all terms are now the amplitudes of the perturbations and quantities of the static background, we can delete all reference to subscripts zero. Furthermore, let us define a new variable $\zeta \equiv \xi/r$, so that Eqs. (\ref{17}) and (\ref{20}) can be rewritten as two first-order time-independent equations
\begin{eqnarray}
\frac{d\zeta}{dr} &=& -\frac{1}{r}\left( 3\zeta + \frac{\Delta p_r}{\gamma p_r} + \frac{2\sigma\zeta}{\epsilon + p_r} \right) + \frac{d\psi}{dr}\zeta ,  \label{21}   \\
\frac{d(\Delta p_r)}{dr} &=&\ \zeta\left\lbrace \frac{\omega^2}{c^2}e^{2(\lambda - \psi)}(\epsilon + p_r)r - 4\frac{dp_r}{dr} \right.  \nonumber  \\
&&\left. - \kappa (\epsilon + p_r)e^{2\lambda}rp_r + r(\epsilon + p_r)\left(\frac{d\psi}{dr}\right)^2 \right.  \nonumber \\
&&\left. + 2\sigma\left(\frac{4}{r} + \frac{d\psi}{dr} \right) + 2\frac{d\sigma}{dr} \right\rbrace + 2\sigma\frac{d\zeta}{dr}   \nonumber \\
&& - \Delta p_r \left[ \frac{d\psi}{dr} + \frac{\kappa}{2}(\epsilon + p_r)re^{2\lambda} \right] + \frac{2}{r}\delta\sigma ,  \label{22}  \qquad
\end{eqnarray}
where the specific form of the perturbation $\delta\sigma$ depends on the anisotropy ansatz that we are going to use afterwards. These equations govern the adiabatic radial oscillations in the interior of an anisotropic spherical star, and we highlight that such equations differ considerably from those obtained by the authors in Ref. \cite{ArbanilMalheiro}. We have a Sturm-Liouville type problem for determining the radial oscillation modes with eigenvalues $\omega_0^2 < \omega_1^2 \cdots < \omega_n^2 < \cdots$, where $n$ stands for the number of nodes inside the anisotropic stellar fluid. It is evident that when $\sigma = 0$, the above system of equations assumes the Gondek's form for isotropic fluids \cite{Gondek, FloresL2010, Pretel2020}.

To numerically solve Eqs. (\ref{21}) and (\ref{22}), it becomes necessary to specify some physically meaningful boundary conditions. Analogous to a vibrating string that is fixed at its endpoints, the radial pulsations in the inner region of a star occur between its center and the surface. Indeed, since Eq. (\ref{21}) has a singularity at the origin, it is required that as $r \rightarrow 0$ the coefficient of $1/r$ term must vanish, namely
\begin{equation}\label{23}
    \Delta p_r = -\frac{2\sigma\zeta}{\epsilon + p_r}\gamma p_r -3\gamma\zeta p_r  \qquad \  {\rm as}  \qquad \  r\rightarrow 0 .
\end{equation}
Meanwhile, at the stellar surface where $p_r(R) = 0$, the appropriate boundary condition is that Lagrangian perturbation of the radial pressure vanishes, this is, 
\begin{equation}\label{24}
    \Delta p_r = 0  \qquad \  {\rm as}  \qquad \  r\rightarrow R .
\end{equation}

Notice that the Lagrangian displacement must vanish at the center due to spherical symmetry, that is $\xi(0)=0$. Nevertheless, our equations are in terms of $\zeta$, so a particularly simple approach that is often adopted is to normalize the eigenfunctions so that $\zeta(0) =1$ at the center. In addition, we point out that in the treatment carried out by Misner et al. \cite{MisnerTW1973} for isotropic configurations, the radial oscillations are described by a second-order ordinary differential equation in the ``renormalized displacement function'' given by $\eta \equiv r^2\xi e^{-\psi}$. In this regard, for anisotropic fluids we obtain the following differential equation governing the adiabatic radial pulsations
\begin{eqnarray}\label{25}
&&\frac{d}{dr}\left[ \mathcal{P}\frac{d\eta}{dr} + \mathcal{P}\frac{2\sigma\eta}{r\gamma p_r}\left( \frac{\gamma p_r}{\epsilon + p_r} + 1 \right) \right]  + \left[ \mathcal{Q} + \frac{\omega^2}{c^2}\mathcal{W} \right]\eta = 0 ,  \nonumber  \\
\end{eqnarray}
where
\begin{eqnarray}
\mathcal{P} &\equiv & \frac{\gamma p_r}{r^2}e^{3\psi + \lambda} ,  \label{26}  \\
\mathcal{W} &\equiv & \frac{\epsilon + p_r}{r^2}e^{\psi + 3\lambda} , \label{27}  \\
\mathcal{Q} &\equiv & \frac{1}{r^2}\left[ \frac{p_r'^2}{\epsilon + p_r} - \frac{4p_r'}{r} - \kappa(\epsilon + p_r)p_te^{2\lambda}   \right. \nonumber  \\ 
&&\left. - \frac{4\sigma}{r(\epsilon + p_r)}\left( p_r' - \frac{\sigma}{r} \right) + \frac{8\sigma}{r^2} + \frac{2}{r}\frac{\delta\sigma}{\xi} \right]e^{3\psi + \lambda} .   \label{28}  \qquad \
\end{eqnarray}
Eq. (\ref{25}) leads to a self-adjoint eigenvalue problem for the oscillation frequencies. Indeed, when the anisotropy vanishes (i.e., $p_t = p_r$), such equation is reduced to the form given in Refs. \cite{MisnerTW1973, KokkotasR2001}.

\subsection{Equation of state and anisotropy ansatz}

The conventional way to tackle the problem of anisotropic configurations is by specifying a barotropic EoS for radial pressure, i.e. $p_r = p_r(\rho)$, and additionally an anisotropy function $\sigma \equiv p_t - p_r$ must be assigned. Here, we are going to use a well known EoS and four different functions for $\sigma$.

Based on the SLy effective nucleon-nucleon interaction, Douchin and Haensel \cite{DouchinHaensel2001} calculated an unified equation of state (the so-called SLy EoS) that covers three main regions of neutron-star interior: outer crust, inner crust and liquid core (consisting of neutrons, protons, electrons and negative muons). Such an EoS is compatible with the constraints of the gravitational-wave event GW170817 (observed by the LIGO-Virgo detectors \cite{Abbott1}), and the analytical parameterization of pressure as a function of density for non-rotating stars is as follows \cite{HaenselPotekhin2004}
\begin{eqnarray}\label{29}
\mathcal{A}(y) &=& \frac{a_1 + a_2y + a_3y^3}{1 + a_4y}K_0(a_5(y - a_6))  \nonumber  \\
&& + (a_7 + a_8y)K_0(a_9(a_{10} - y))   \nonumber  \\
&& + (a_{11} + a_{12}y)K_0(a_{13}(a_{14} - y))  \nonumber  \\
&& + (a_{15} + a_{16}y)K_0(a_{17}(a_{18} - y)) ,  \qquad
\end{eqnarray}
where $\mathcal{A} \equiv \log(p_r/ \rm dyn\ cm^{-2})$, $y \equiv \log(\rho/ \rm g\ cm^{-3})$, and $K_0(x) \equiv 1/(e^x +1)$. The fitting parameters of this expression $a_i$ can be found in Ref. \citep{HaenselPotekhin2004}. Furthermore, below we specify the anisotropy functions provided in the literature to model anisotropic matter at high densities:

\subsubsection{Quasi-local ansatz} Horvat et al. \cite{Horvat2011} suggested an anisotropy ansatz as being a bilinear function in the radial pressure and compactness, namely
\begin{equation}\label{30}
\sigma \equiv \beta_{\rm H} p_r \mu ,
\end{equation}
where $\beta_{\rm H}$ is a dimensionless parameter that measures the degree of anisotropy within the star, and $\mu(r) \equiv 2Gm/c^2r$ is known as compactness. The advantage of this ansatz is that the fluid becomes isotropic at the stellar center since $\mu \sim r^2$ when $r \rightarrow 0$. On the other hand, (\ref{30}) is important only for relativistic configurations which is in agreement with the assumption that the anisotropy may arise at high densities. A similar ansatz for the anisotropy measure was used to describe gravastar models in Ref. \cite{DeBenedictis2006}. Moreover, in the calculations it is common to assume $-2 \leq \beta_{\rm H} \leq 2$ \cite{Horvat2011, DonevaYazadjiev, Folomeev2015, Silva2015, Folomeev2018}.

According to Eq. (\ref{8}), the Eulerian perturbation for the anisotropy function (\ref{30}), can be written as
\begin{eqnarray}\label{31}
\delta\sigma &=& \beta_{\rm H} (1 - e^{-2\lambda})(\Delta p_r - r\zeta p_r')  \nonumber  \\
&& - \beta_{\rm H}\kappa p_r(\epsilon + p_r) r^2\zeta .
\end{eqnarray}

\subsubsection{Bowers-Liang ansatz} Another relation for anisotropic models was proposed by Bowers and Liang \cite{BowersLiang}, given by
\begin{equation}\label{32}
\sigma \equiv \beta_{\rm BL}\frac{G}{c^4}(\epsilon + p_r)(\epsilon + 3p_r)\frac{r^2}{1-\mu} ,
\end{equation}
where the anisotropy factor depends nonlinearly on the radial pressure and energy density. The anisotropy vanishes at the origin in order to yield regular solutions, and it is (in part) gravitationally induced since $1- \mu = e^{-2\lambda}$. Here the literature offers a similar range for the free parameter $\beta_{\rm BL}$ as in the first model mentioned above (see Refs. \cite{Silva2015, Folomeev2018, Biswas2019}).

In this case, $\delta\sigma$ assumes the form
\begin{eqnarray}\label{33}
\delta\sigma &=& \frac{2\beta_{\rm BL} G}{c^4} r^2e^{2\lambda} \left[ (2\epsilon + 3p_r)(\Delta p_r - r\zeta p_r')  \right.  \nonumber  \\  
&&\left. + (\epsilon + p_r)(\epsilon + 3p_r)(\zeta + \delta\lambda) + (\epsilon + 2p_r)\delta\epsilon \right] .   \qquad
\end{eqnarray}

\subsubsection{Herrera-Barreto ansatz} An ansatz that was initially studied in Ref. \cite{CHEW1981} in order to find a family of non-isotropic configurations from any isotropic model, and later summarized by Herrera and Barreto \cite{HerreraBarreto}, is the following
\begin{equation}\label{34}
\sigma \equiv \frac{(h-1)r}{2h} \frac{dp_r}{dr} ,
\end{equation}
where $h$ is a constant throughout the sphere, and for $h=1$ we recover the isotropic case. Here we are going to define $\beta_{\rm HB} \equiv (h-1)/2h$ so that $\sigma = \beta_{\rm HB} rp_r'$. The possibility (\ref{34}), like the other ansatze, guarantees that the anisotropy must vanish at the center of symmetry of the fluid. In this case $\beta_{\rm HB}$ cannot be positive in order for the tangential pressure to be always positive throughout the stellar interior. We must point out that physically relevant solutions correspond to $p_r, p_t \geq 0$ for $r \leq R$.

The Eulerian perturbation for the ansatz (\ref{34}) is given by
\begin{equation}\label{35}
\delta\sigma = \beta_{\rm HB} r\zeta p_r' + \beta_{\rm HB} r\frac{d}{dr}(\Delta p_r - r\zeta p_r') .
\end{equation}

\subsubsection{Covariant ansatz} Finally we will consider an additional ansatz that has recently been introduced by Raposo et al. \cite{Raposo2019} in order to study the dynamical properties of anisotropic self-gravitating fluids in a covariant framework, that is, $\sigma \equiv -\mathcal{C}f(\epsilon)k^\mu \nabla_\mu p_r = -\mathcal{C}f(\epsilon)e^{-\lambda}p_r'$, where $\mathcal{C}$ is a free constant that measures the deviation from isotropy. For simplicity, we are going to consider $f(\epsilon) = \epsilon$, and since we are using physical units, we have
\begin{equation}\label{36}
\sigma = \beta_{\rm R}\frac{G}{c^4}\epsilon e^{-\lambda}p_r' ,
\end{equation}
where now the free parameter is $\beta_{\rm R} \leq 0$ and, unlike previous models, it has units of cubic meters. At the stellar origin the fluid becomes isotropic since the radial pressure is maximum at this point, and at the surface both pressures vanish.

For this anisotropic model, $\delta\sigma$ takes the following form
\begin{equation}\label{37}
\delta\sigma = \frac{\beta_{\rm R}G}{c^4} e^{-\lambda}\left[ (\delta\epsilon - \epsilon \delta\lambda)p_r' + \epsilon \frac{d}{dr}(\Delta p_r - r\zeta p_r') \right] .  
\end{equation}

%--------------------------------------------------------------

\section{Non-adiabatic gravitational collapse}\label{Sec3}

The goal in this section is to study the dynamical evolution of unstable anisotropic neutron stars whose ultimate fate is the formation of an event horizon, and to describe the formation of black holes it is necessary to use models that involve non-ideal fluids \cite{Terno2019}. The gravitational collapse is a highly dissipative  phenomenon in which massless particles (photons and neutrinos) carry thermal energy for exterior spacetime \cite{HerreraSantos2004, Mitra}. In this respect, we are going to deal with the problem following the standard procedure, namely, the spherical hypersurface $\Sigma$ of the collapsing star divides the spacetime into two different regions where each one is described by a particular matter-energy distribution.

\subsection{Interior spacetime and generalized TOV equations}

We model the collapsing configuration by means of a locally anisotropic fluid, bounded by $\Sigma$, and that undergoes dissipation in the form of heat flow. Accordingly, in the diffusion approximation, the energy-momentum tensor is given by \cite{HerreraSantos2004}
\begin{eqnarray}\label{38}
T_{\mu\nu}^- &=& (\epsilon + p_t)u_\mu u_\nu + p_tg_{\mu\nu} + (p_r - p_t)k_\mu k_\nu  \nonumber  \\
&& + q_\mu u_\nu + q_\nu u_\mu ,
\end{eqnarray}
where, as in the static background, $\epsilon$ represents the energy density, $p_r$ the radial pressure and $p_t$ the tangential pressure. The four-velocity $u^\mu$, the heat flux $q^\mu$ and the unit four-vector along the radial direction $k^\mu$, must satisfy the following relations
\begin{equation}\label{39}
u_\mu u^\mu = -1,  \qquad   k_\mu k^\mu = 1,  \qquad  q_\mu u^\mu = 0 .
\end{equation}

In order to introduce a time dependence on the metric functions such that under a certain limit we can recover the static case, we assume that the geometry of the interior spacetime is described by the following spherically symmetric, shear-free line element \cite{Pretel2020}
\begin{eqnarray}\label{40}
ds_-^2 = g_{\mu\nu}^- dx^\mu_- dx^\nu_- &=& -e^{2\psi(r)}(dx^0_-)^2 + e^{2\lambda(r)}f(x^0_-)dr^2 \nonumber \quad  \\
&& + r^2f(x^0_-)d\Omega^2 ,
\end{eqnarray}
where $x^\mu_- = (ct, r, \theta, \phi)$ are the coordinates in the interior manifold and $df/dx_-^0 <0$ in the case of collapsing configurations. From relations (\ref{39}) and by using comoving coordinates, we get
\begin{equation}\label{41}
u^\mu = e^{-\psi}\delta_0^\mu ,  \qquad   k^\mu = \frac{e^{-\lambda}}{\sqrt{f}}\delta_r^\mu ,  \qquad   q^\mu = \frac{q}{c}\delta_r^\mu ,
\end{equation}
being $q = q(x^0_-, r)$ the rate of energy flow per unit area along the radial coordinate. In the static limit $f(x^0_-) \rightarrow 1$ we recover Eq. (\ref{1}) which describes the initially static anisotropic star.

The explicit form of the Einstein field equations for the energy-momentum tensor (\ref{38}) and metric (\ref{40}), is given by
\begin{eqnarray}
G_{00}^- &=& -\frac{e^{2\psi}}{r^2e^{2\lambda}f}( 1- e^{2\lambda} - 2r\lambda' ) + \frac{3}{4}\frac{\dot{f}^2}{f^2} = \kappa\epsilon e^{2\psi} ,  \label{42}  \\
G_{rr}^- &=& \frac{1- e^{2\lambda} + 2r\psi'}{r^2} - \frac{e^{2\lambda}}{e^{2\psi}}\left( \ddot{f} - \frac{\dot{f}^2}{4f} \right) = \kappa p_r e^{2\lambda}f ,  \qquad  \label{43}   \\
G_{\theta\theta}^- &=& \frac{r^2}{e^{2\lambda}}\left( \psi'^2 + \psi'' - \psi'\lambda' + \frac{\psi'}{r} - \frac{\lambda'}{r} \right)  \nonumber  \\
&& - \frac{r^2}{e^{2\psi}}\left( \ddot{f} - \frac{\dot{f}^2}{4f} \right) = \kappa p_t r^2f ,  \label{44}  \\
G_{0r}^- &=& \psi'\frac{\dot{f}}{f} = -\kappa\frac{q}{c}e^{\psi + 2\lambda}f ,  \label{45}
\end{eqnarray}
where the dot and the prime denote differentiation with respect to $x^0_-$ and $r$, respectively. The mass entrapped within the radius $r$ and at time $t$, is given by the following expression \cite{HerreraSantos2004, Cahill1970}
\begin{eqnarray}\label{46}
m(t,r) &=& \frac{c^2r\sqrt{f}}{2G}R^{\phi}_{\ \theta\phi\theta}  \nonumber  \\
&=& \frac{c^2r\sqrt{f}}{2G}\left[ 1 - \frac{1}{e^{2\lambda}} + \frac{r^2}{4e^{2\psi}}\frac{\dot{f}^2}{f} \right] .
\end{eqnarray}

As a consequence, the generalized TOV equations within a non-adiabatic context are derived from the four-divergence of the energy-momentum tensor (\ref{38}) along with Eq. (\ref{43}), that is,
\begin{eqnarray}
\frac{\partial p_r}{\partial r} &=& -\frac{\epsilon + p_r}{c^2}\left[ \frac{Gm}{r^2\sqrt{f}} + \frac{4\pi G}{c^2}rp_rf + \frac{c^2r}{2e^{2\psi}}\left( \ddot{f} - \frac{\dot{f}^2}{2f} \right) \right]  \nonumber  \\
&& \times \left[ 1- \frac{2Gm}{c^2r\sqrt{f}} + \frac{r^2\dot{f}^2}{4e^{2\psi}f} \right]^{-1} + \frac{2}{r}\sigma   \nonumber  \\
&& - \frac{f}{ce^\psi}\left[ \dot{q} + \frac{5q\dot{f}}{2f} \right] \left[ 1- \frac{2Gm}{c^2r\sqrt{f}} + \frac{r^2\dot{f}^2}{4e^{2\psi}f} \right]^{-1} ,  \label{47}  \\
\frac{d\psi}{dr} &=& -\frac{1}{\epsilon + p_r}\frac{\partial p_r}{\partial r} + \frac{2\sigma}{r(\epsilon + p_r)}  \nonumber  \\
&& - \frac{e^{2\lambda - \psi}f}{c(\epsilon + p_r)}\left( \dot{q} + \frac{5q\dot{f}}{2f} \right) ,  \label{48}
\end{eqnarray}
where it should be noted that now the energy density, radial pressure, heat flux, anisotropy ansatz, and mass function depend on both $t$ and $r$, whereas $\psi = \psi(r)$, $\lambda = \lambda(r)$ and $f = f(t)$. It becomes apparent that at the static limit such equations are reduced to those already known in (\ref{6}) and (\ref{7}), respectively. Furthermore, we remark that Eq. (\ref{46}) is the dissipative analogue of the relation (\ref{8}).

\subsection{Junction conditions on the stellar surface}

Since the star is radiating energy, the exterior spacetime is described by the Vaidya metric \cite{Vaidya, Mkenyeleye}, given as follows
\begin{eqnarray}\label{49}
ds_+^2 = g_{\mu\nu}^+ dx^\mu_+ dx^\nu_+ &=& -\left[ 1 - \frac{2Gm(\upsilon)}{c^2\chi} \right]c^2d\upsilon^2  \nonumber \\
&& + 2e c d\upsilon d\chi + \chi^2d\Omega^2 ,  \qquad \ \
\end{eqnarray}
where $x_+^\mu = (c\upsilon, \chi, \theta, \phi)$, $m(\upsilon)$ is the mass function that depends on the retarded time $\upsilon$, and $e = \pm 1$ describes the incoming (outgoing) flux of radiation around the source of gravitational field. In our collapse model the radiation is expelled into outer region so that $dm/d\upsilon \leq 0$.

Since the collapsing star is described by two spacetime regions with distinct geometric properties, it becomes necessary to invoke the junction conditions on $\Sigma$ established by Israel \cite{Israel1, Israel2}. Such conditions require continuity of the line element and extrinsic curvature through the hypersurface, namely
\begin{equation}\label{50}
(ds^2_-)_\Sigma = (ds^2_+)_\Sigma = ds^2_\Sigma ,  \ \quad   (K_{ij}^-)_\Sigma = (K_{ij}^+)_\Sigma ,  
\end{equation}
where the intrinsic metric to $\Sigma$ is given by
\begin{equation}\label{51}
ds^2_\Sigma = g_{ij}d\varsigma^i d\varsigma^j = -c^2d\tau^2 + \mathcal{R}^2(\tau)d\Omega^2 . 
\end{equation}

The extrinsic curvature tensor is defined by
\begin{equation}\label{52}
K_{ij}^\pm = -n_\mu^\pm \frac{\partial^2 x_\pm^\mu}{\partial\varsigma^i \partial\varsigma^j} - n_\mu^\pm \Gamma_{\alpha\beta}^\mu \frac{\partial x_\pm^\alpha}{\partial \varsigma^i}\frac{\partial x_\pm^\beta}{\partial \varsigma^j} ,
\end{equation}
where $x_\pm^\mu$ are the coordinates of the exterior and interior  spacetime, $\varsigma^i = (c\tau, \theta, \phi)$ are the coordinates that define the comoving timelike hypersurface, and $n_\pm^\mu$ are the unit normal vectors to $\Sigma$ which have already been calculated by Santos \cite{Santos1985}. The non-vanishing extrinsic curvature components $K_{ij}^\pm$ are given explicitly in Appendix \ref{ApendixA}. As a result, the junction conditions (\ref{50}) imply that
\begin{eqnarray}
    \chi_\Sigma &=& \left[r\sqrt{f}\right]_\Sigma = \mathcal{R} ,  \label{53}  \\
    m_\Sigma &=& \frac{c^2R\sqrt{f}}{2G}\left[ 1 + \frac{r^2}{4e^{2\psi}}\frac{\dot{f}^2}{f} - \frac{1}{e^{2\lambda}} \right]_\Sigma ,  \label{54}   \\
    z_\Sigma &=& \left[ \frac{d\upsilon}{d\tau} \right]_\Sigma - 1 = \left[ \frac{1}{e^\lambda} + \frac{r}{2e^\psi}\frac{\dot{f}}{\sqrt{f}} \right]_\Sigma^{-1} - 1 ,  \quad   \label{55}  \\
    p_{r, \Sigma} &=& \left[ \frac{q}{c}e^{\lambda}\sqrt{f} \right]_\Sigma ,   \label{56} 
\end{eqnarray}
with $z_\Sigma$ being the boundary redshift of the radial radiation emitted by the non-adiabatic sphere. Eq. (\ref{53}) is the equality of the proper radii as measured from the perimeter of $\Sigma$. The expression (\ref{54}) is a measure of the total mass of the star as it collapses, and Eq. (\ref{56}) indicates that the radial pressure at the surface of the star is different from zero unless the heat flow vanishes. It is evident that in the static limit $m_\Sigma$ corresponds to the total mass of the initial Schwarzschild configuration $M$, and the redshift is reduced to $z_\Sigma = e^{\lambda(R)} - 1$, namely, the gravitational redshift of light emitted at the surface of the initially static neutron star.

\subsection{Evolution quantities}

The fact that the radial pressure does not vanish at the surface leads to an additional differential equation that allows us to fix the time evolution of our model. By taking into account Eqs. (\ref{43}) and (\ref{45}) into (\ref{56}), we obtain
\begin{equation}\label{57}
    \frac{d^2f}{dt^2} - \frac{1}{4f}\left( \frac{df}{dt} \right)^2 - \frac{GM}{cR^2}\frac{1}{\sqrt{f}}\frac{df}{dt} = 0 ,
\end{equation}
which can be integrated to generate the following equation
\begin{equation}\label{58}
    \frac{df}{dt} = \frac{4GM}{cR^2}\left[ \sqrt{f} - f^{1/4} \right] ,  
\end{equation}
where the integration constant has been determined by applying the static limit. Then the solution of Eq. (\ref{58}) is given by
\begin{equation}\label{59}
    t = \frac{cR^2}{2GM}\left[ \sqrt{f} + 2f^{1/4} + 2\ln\left(1 - f^{1/4}\right) \right] .
\end{equation}

For an observer at rest at infinity, the redshift (\ref{55}) diverges at the time of formation of an event horizon. This means that a black hole has been formed as outcome of the gravitational collapse of an unstable anisotropic star when
\begin{equation}\label{60}
f_{bh} = \left[ \frac{2GM}{c^2R} \right]^4 .
\end{equation}
For systems describing gravitational collapse we must have $0 < f \leq 1$, i.e. the time function $f$ decrease monotonically from  $f = 1$ to $f = f_{bh}$. In other words, the time goes from $t = -\infty$ (when the model is static) to $t = t_{bh}$ (when the star becomes a black hole), but a time displacement can be done without loss of generality. Consequently, one obtains from Eq. (\ref{54}) the mass of the formed black hole, which reads:
\begin{equation}\label{61}
m_{bh} = \frac{2GM^2}{c^2R} .
\end{equation} 

Notice that Eq. (\ref{59}) provides $t$ as a function of $f$, nonetheless, is more useful to obtain $f(t)$ in order to analyze the dynamical quantities as a function of time as a stellar configuration collapses. Thus, it is convenient to numerically solve Eq. (\ref{57}) as a final value problem by specifying a value of $f(t)$ and $df(t)/dt$ at time $t = t_{bh}$, where the two final conditions are established through Eqs. (\ref{58})-(\ref{60}). The relevant physical quantities during the collapse such as energy density, radial pressure, tangential pressure and heat flow, are given by
\begin{eqnarray}
\epsilon(t, r) &=&  -\frac{1 - e^{2\lambda} - 2r\lambda'}{\kappa r^2e^{2\lambda}f} + \frac{12a^2}{\kappa e^{2\psi}}\left[ \frac{\sqrt{f} - f^{1/4}}{f} \right]^2  \nonumber  \\
&=& \frac{2}{\kappa r^2f}\left[ \frac{1 + r\lambda'}{e^{2\lambda}} + \frac{3Gm}{c^2r\sqrt{f}} -1 \right] ,   \label{62}  \\
p_r(t, r) &=& \ \frac{1 - e^{2\lambda} + 2r\psi'}{\kappa r^2e^{2\lambda}f} + \frac{4a^2}{\kappa e^{2\psi}}\left[ \frac{f^{-1/4} - 1}{f} \right] ,  \qquad \  \label{63}  \\
p_t(t, r) &=& \frac{1}{\kappa e^{2\lambda}f}\left[ \psi'^2 + \psi'' - \psi'\lambda' + \frac{1}{r}(\psi'-\lambda') \right]  \nonumber  \\
&& + \frac{4a^2}{\kappa e^{2\psi}}\left[ \frac{f^{-1/4} - 1}{f} \right] ,  \label{64}  \\
q(t, r) &=& -\frac{4ac\psi'}{\kappa e^{\psi +2\lambda}}\left[ \frac{\sqrt{f} - f^{1/4}}{f^2} \right] ,    \label{65}
\end{eqnarray}
where $a \equiv GM/c^2R^2$, and the anisotropy factor takes the following form
\begin{eqnarray}\label{66}
\sigma(t, r) &=& \frac{1}{\kappa e^{2\lambda}f}\bigg[ \psi'^2 + \psi'' - \psi'\lambda'   \nonumber \\
&& \hspace{1.7cm} \left. + \frac{1}{r^2}(e^{2\lambda} - 1 - r\psi' - r\lambda') \right] .  \qquad
\end{eqnarray}

Finally, a kinematic quantity that provides information about the rate of expansion of the fluid sphere is given by the four-divergence of the four-velocity \cite{Rezzolla, HerreraSantos2004}
\begin{equation}\label{67}
\Theta = c\nabla_\mu u^\mu = \frac{6ac}{e^\psi}\left[ \frac{\sqrt{f} - f^{1/4}}{f} \right] ,
\end{equation}
that is, the expansion scalar and whose action is to change the volume of the spherical star but it preserves the principal axes. In fact, when $f \rightarrow 1$, the heat flow and the expansion scalar vanish and, therefore, the fluid becomes perfect. The other physical quantities are reduced to those already known in the static background.

\subsection{Energy conditions}

The dissipative anisotropic fluid must satisfy the energy conditions throughout the gravitational collapse in order for it to be physically acceptable. This means that the energy-momentum tensor (\ref{38}) has to be diagonalized through equation $\vert T_{\mu\nu}^- - \Upsilon g_{\mu\nu}^- \vert = 0$, so that the eigenvalues $\Upsilon$ take the explicit form 
\begin{eqnarray}
    \Upsilon_0 &=& -\frac{1}{2}(\epsilon - p_r + \Delta) , \label{68}   \\
    \Upsilon_1 &=& -\frac{1}{2}(\epsilon - p_r - \Delta) , \label{69}   \\
    \Upsilon_2 &=& \Upsilon_3 = p_t , \label{70}
\end{eqnarray}
where we defined $\Delta \equiv \sqrt{ (\epsilon + p_r)^2 - 4\tilde{q}^2 }$ and $\tilde{q} \equiv \frac{q}{c}e^\lambda \sqrt{f}$. Thus, the following energy conditions must hold \cite{Kolassis1988} 
\begin{itemize}
    \item[$\star$] \textbf{Weak energy conditions (WEC)} 

    \begin{itemize}
    \item[a)] $-\Upsilon_0 \geq 0$,
    \item[b)] $-\Upsilon_0 + \Upsilon_i \geq 0, \quad   \rm{for} \ i = 1, 2, 3.$
    \end{itemize}
        
    Such inequalities entail that
    \begin{eqnarray} 
        &&\Delta \geq 0,  \label{71}  \\  
        &&\epsilon + \Delta - p_r \geq 0 ,  \label{72}  \\  
        &&\epsilon + \Delta - p_r +2p_t \geq 0 .  \label{73}
    \end{eqnarray}
    
    \item[$\star$] \textbf{Dominant energy conditions (DEC)} 
    
    \begin{itemize}
    \item[a)] $-\Upsilon_0 \geq 0$ , 
    \item[b)] $-\Upsilon_0 + \Upsilon_i \geq 0, \quad   \rm{for} \ i = 1, 2, 3.$ 
    \item[c)] $\Upsilon_0 + \Upsilon_i \leq 0, \quad   \rm{for} \ i = 1, 2, 3.$
    \end{itemize}
       
    The first two inequalities have already been included in weak energy conditions. As regards the third inequality, we obtain
    \begin{eqnarray}
        &&\epsilon - p_r \geq 0 ,  \label{74}  \\  
        &&\epsilon + \Delta - p_r - 2p_t \geq 0 .   \label{75} 
    \end{eqnarray}

    \item[$\star$] \textbf{Strong energy conditions (SEC)}
    \vspace*{0.2cm}
    \begin{itemize}
    \item[a)] $-\Upsilon_0 + \sum_{i=1}^3 \Upsilon_i \geq 0$ ,
    \item[b)] $-\Upsilon_0 + \Upsilon_i \geq 0, \quad   \rm{for} \ i = 1, 2, 3.$
    \end{itemize}
    \vspace*{0.2cm}
    Specifically, the first inequality implies that
    \begin{equation}\label{76}
     \hspace{0.3cm}   \Delta + 2p_t \geq 0 ,
    \end{equation}
    whereas the second inequality has been considered in the other energy conditions.
\end{itemize}

Therefore, to know whether our collapse model is physically acceptable, it is only necessary to verify that the energy conditions (\ref{71}), (\ref{74}), (\ref{75}) and (\ref{76}) are respected.

%--------------------------------------------------------------

\section{Numerical results and discussion}\label{Sec4}

\subsection{Equilibrium configurations and radial oscillations}

\begin{figure*}
 \includegraphics[width=8.4cm]{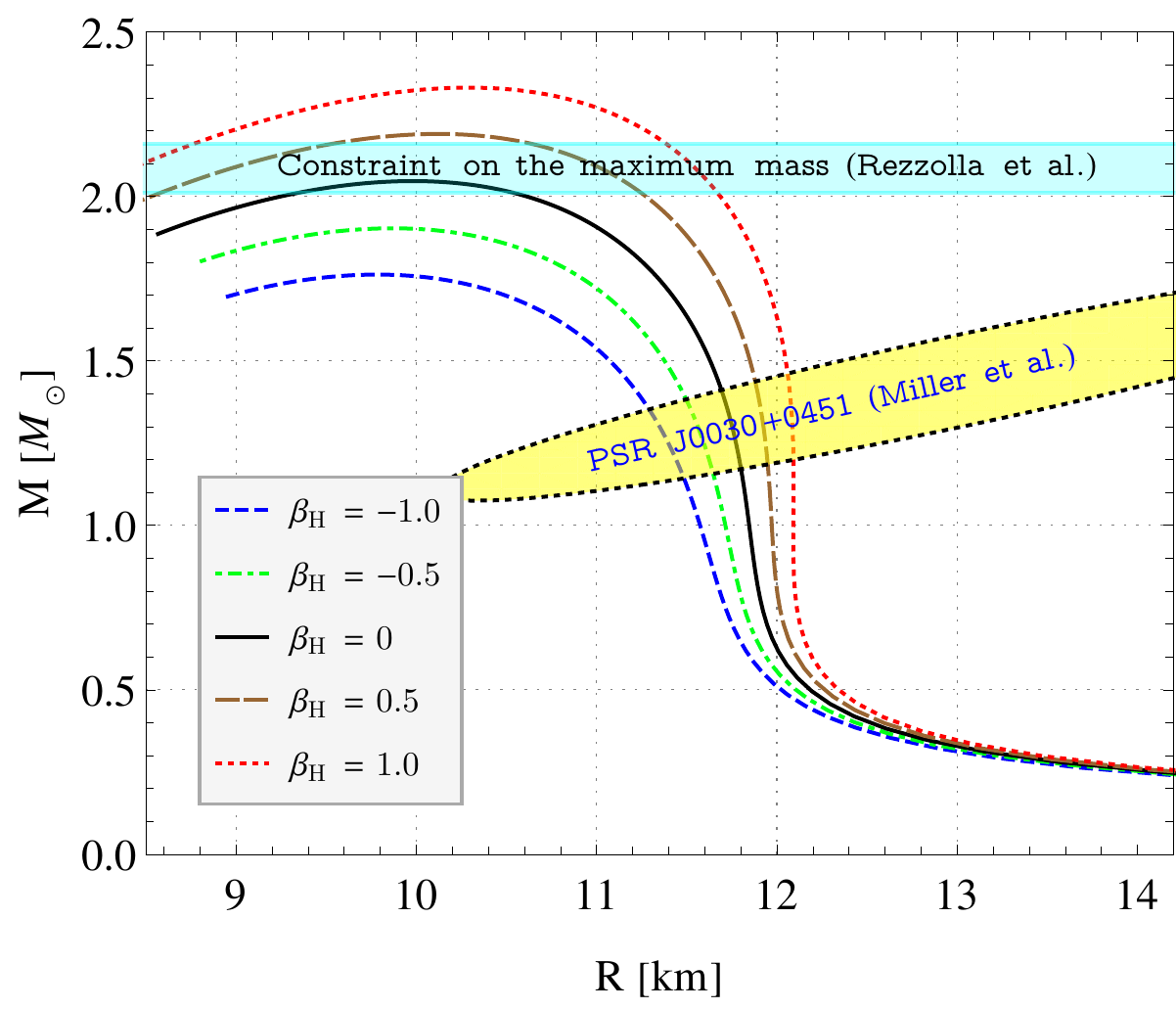} \
 \includegraphics[width=8.4cm]{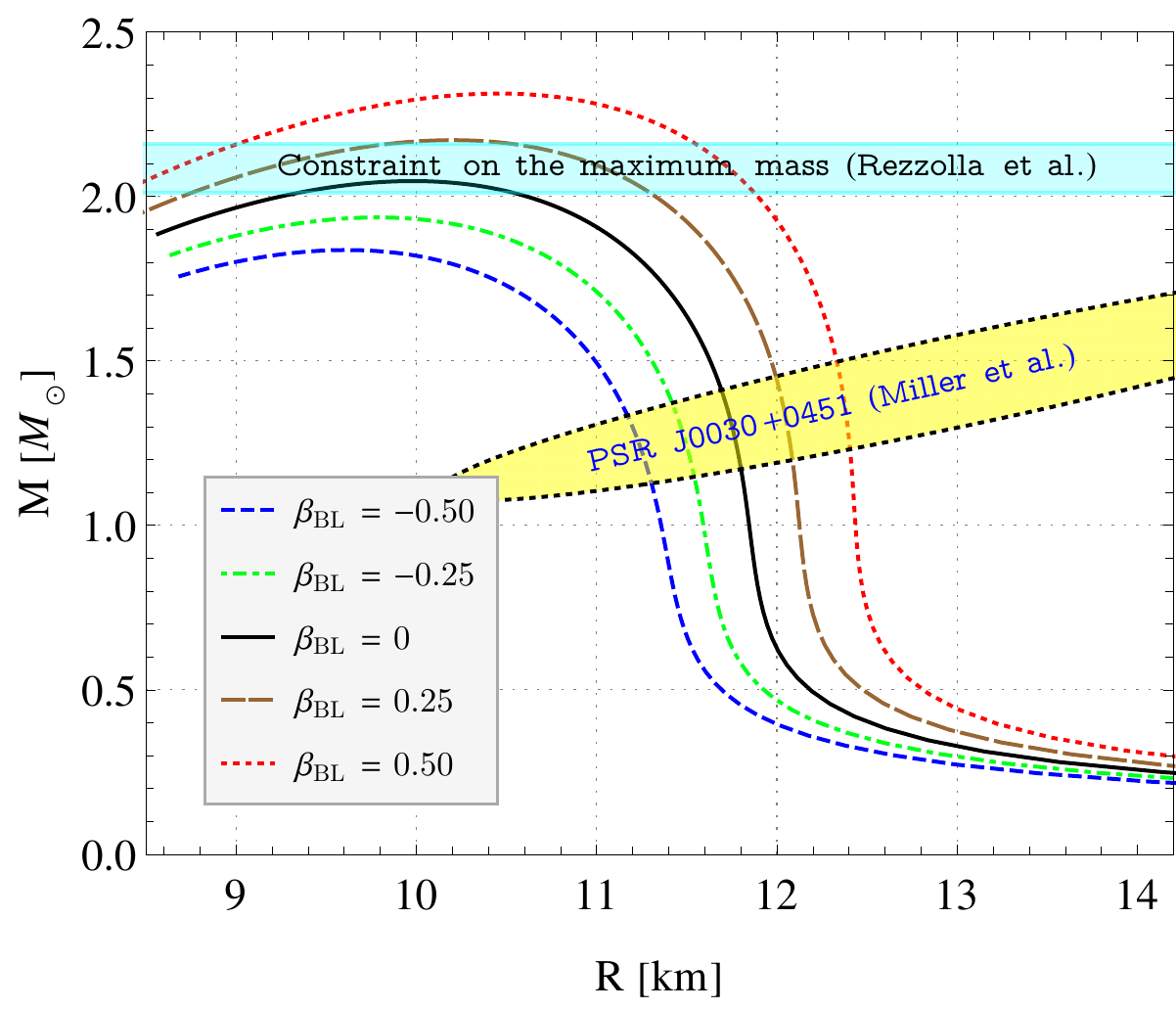}
 \includegraphics[width=8.4cm]{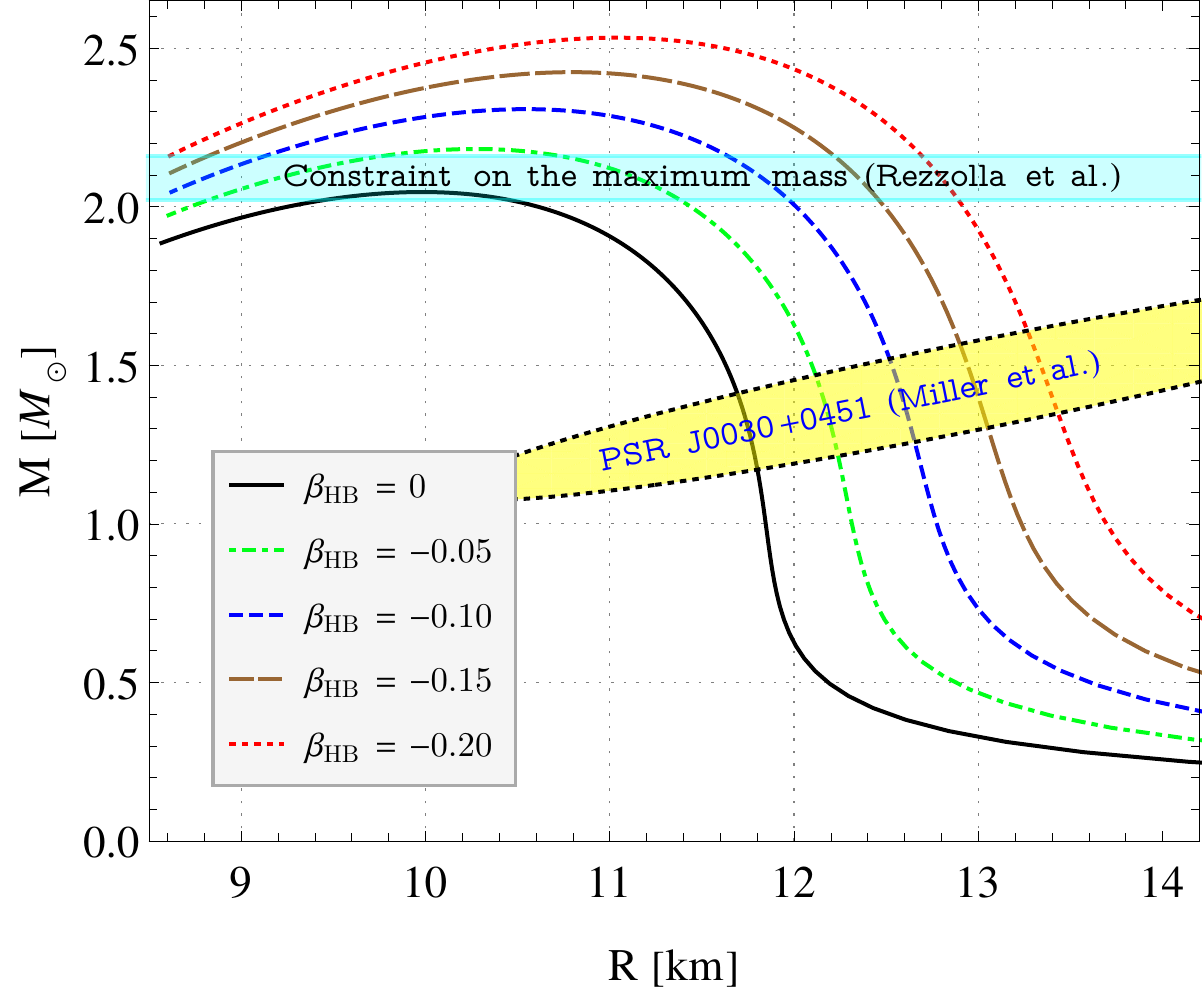} \
 \includegraphics[width=8.4cm]{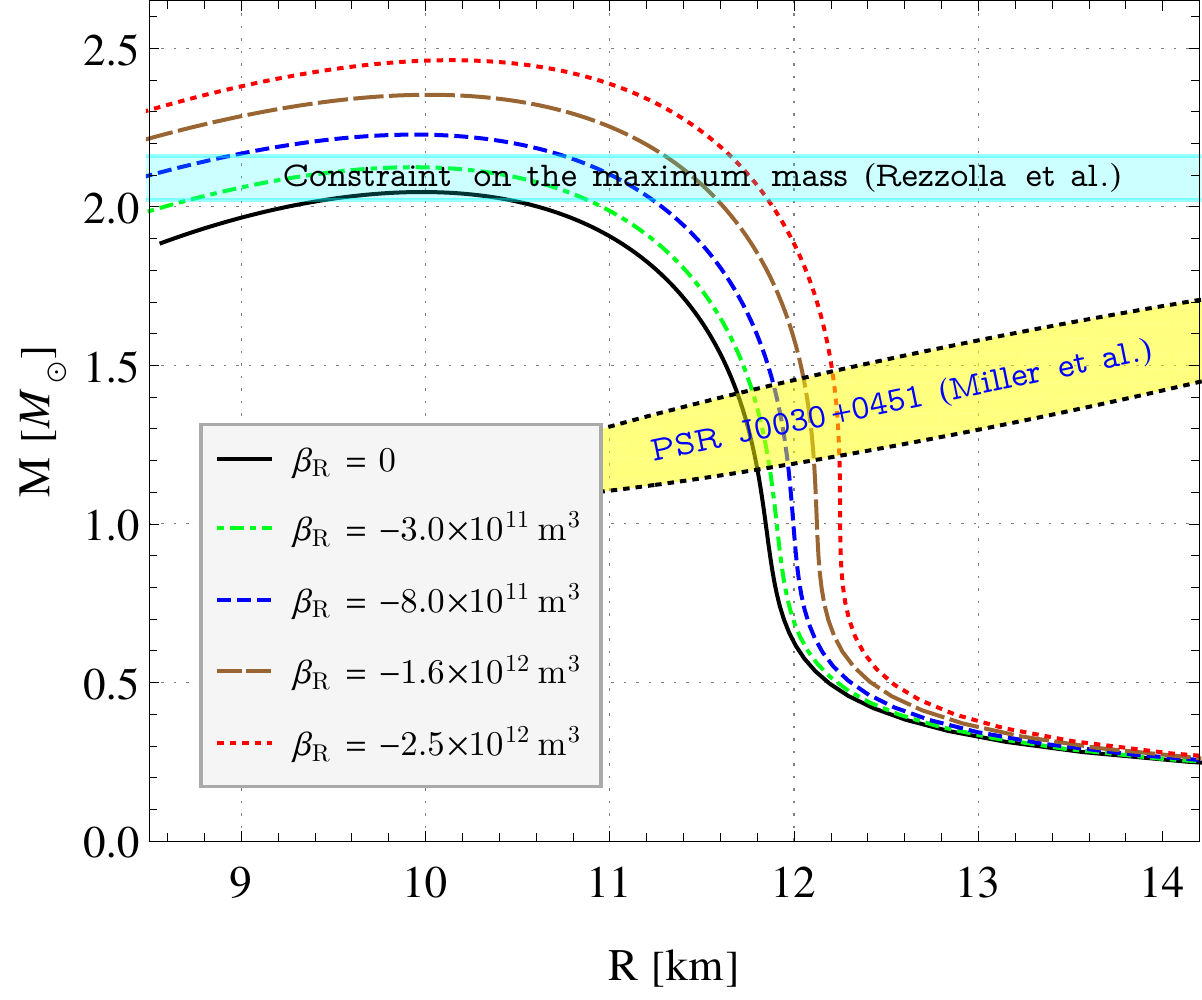}
 \caption{\label{figure1} Mass-radius diagrams for anisotropic neutron stars with SLy EoS (\ref{29}) and for the anisotropy ansatze (\ref{30}) in the upper left panel, (\ref{32}) in the upper right panel, (\ref{34}) in the lower left panel, and (\ref{36}) in the lower right panel. The isotropic case is shown in all plots as a benchmark by a solid black line, and the cigar-shaped yellow region is one of both mass and radius measurements for PSR J0030+0451 from NICER data \cite{Miller2019}. The horizontal narrow band in cyan color stands for the recent restriction of maximum mass of neutron stars as a result of observation of the GW event GW170817 \cite{RezzolaMW2018}.}
\end{figure*}

Given different values of central density and an anisotropy ansatz with EoS (\ref{29}), the background equations (\ref{5})-(\ref{7}) can be numerically solved under the conditions (\ref{9}) to produce a family of anisotropic neutron stars in hydrostatic equilibrium. Figure \ref{figure1} illustrates the landscape of these configurations for each anisotropy ansatz. In the case of the ansatze proposed by Horvat et al. (\ref{30}) and Raposo et al. (\ref{36}), for low enough central densities, the configurations have a mass similar to the isotropic case. Nevertheless, above a certain central density value, the masses deviate considerably from those provided by isotropic solutions. It is evident that depending on the degree of anisotropy within the configurations, the curves in the $M$ versus $R$ plane can move significantly away from the isotropic case.

One of the most interesting approaches to determining the neutron-star matter EoS is through measurements of the masses and radii of these stars. The observations made in the last few years are allowing us to improve our understanding of the properties of cold dense matter and, therefore, to constrain the EoS. In that regard, we expect that the effects generated by anisotropic pressure are also within the constraints obtained by recent observations. Indeed, we consider one of both mass and radius measurements for the millisecond pulsar PSR J0030+0451 from Neutron Star Interior Composition Explorer (NICER) data, which was obtained  by using a Bayesian inference approach to analyze its energy-dependent thermal X-ray waveform \cite{Miller2019}. According to Fig. \ref{figure1}, it is possible to construct anisotropic neutron stars that lie within the region provided by the NICER data. Furthermore, from the observation of the GW event GW170817 there is a recent restriction on the maximum mass of neutron stars \cite{RezzolaMW2018}. To ensure that our results are consistent with such a restriction, bounds on anisotropy parameters must be established. This hints that $\beta_{\rm H} \lesssim 0.40$, $\beta_{\rm BL} \lesssim 0.23$, $\beta_{\rm HB} \gtrsim -0.042$, and $\beta_{\rm R} \gtrsim -4.6 \times 10^{11}\ \rm m^3$.

In table \ref{table1}, we list the mass and radius of the maximum-mass configurations for each anisotropic model with different values of the free parameter. According to the $M(\rho_c)$ method, the first maximum on the $M(\rho_c)$ curve corresponds to a critical central density $\rho_c = \rho_{crit}$ which delimits a family of stars that is stable against gravitational collapse. This means that the unstable branch in the sequence of stars is located after the critical density where $dM/d\rho_c <0$. Due to its simplicity, this condition has been widely used in the literature. However, such a condition is just necessary but not sufficient to determine the limits of stability.

\begin{figure} 
\includegraphics[width=8.2cm]{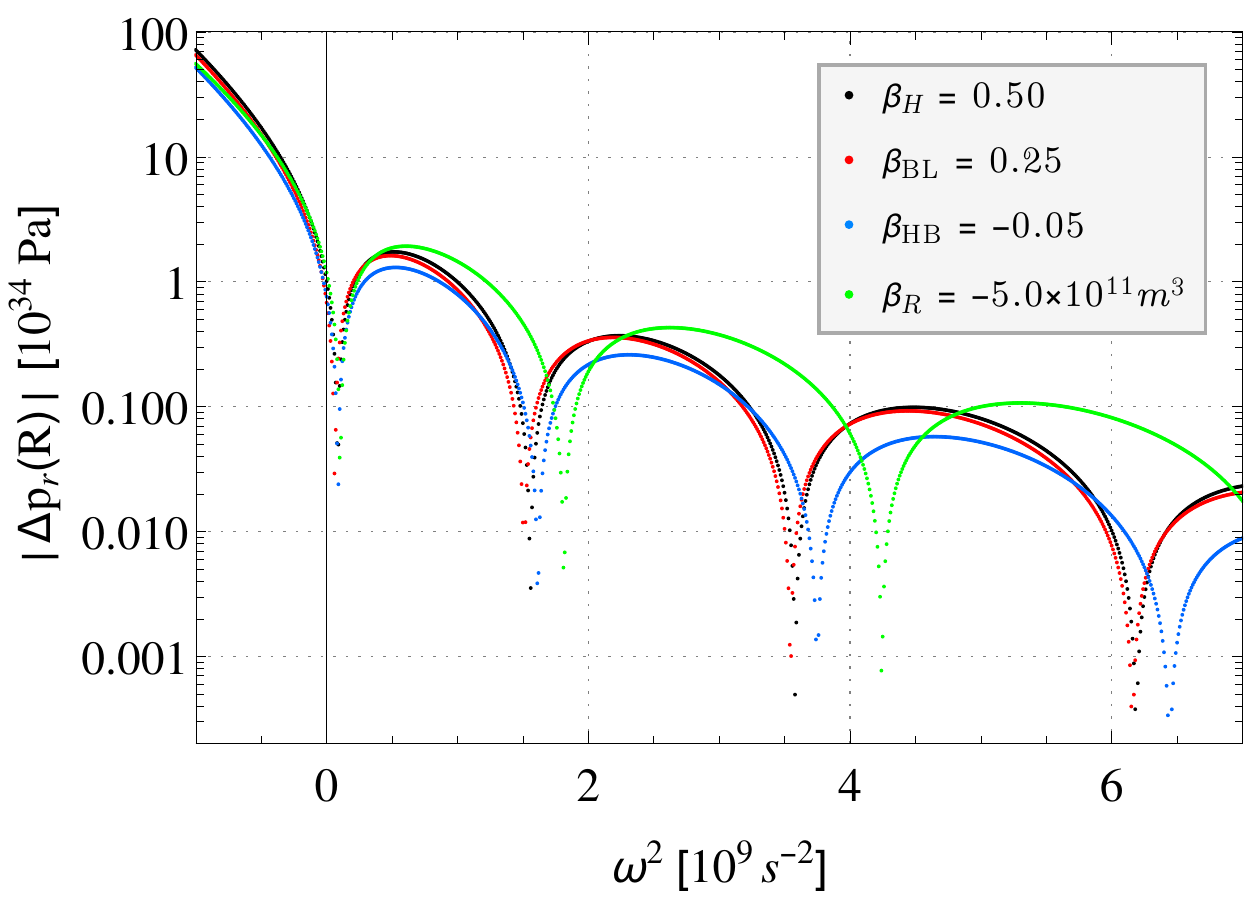}
\caption{Absolute value of the Lagrangian perturbation of the radial pressure at the stellar surface on a logarithmic scale for a set of trial values of $\omega^2$ with a central mass density $\rho_c = 2.0 \times 10^{18}\ \rm{kg}/\rm{m}^3$ and different anisotropy ansatze. The minima in each curve correspond to the correct frequencies of the oscillation modes for equilibrium configurations. Since $\omega_0^2 >0$, the four anisotropic neutron stars are stable against radial oscillations.}
\label{figure2}
\end{figure}

\begin{figure*}
 \includegraphics[width=8.4cm]{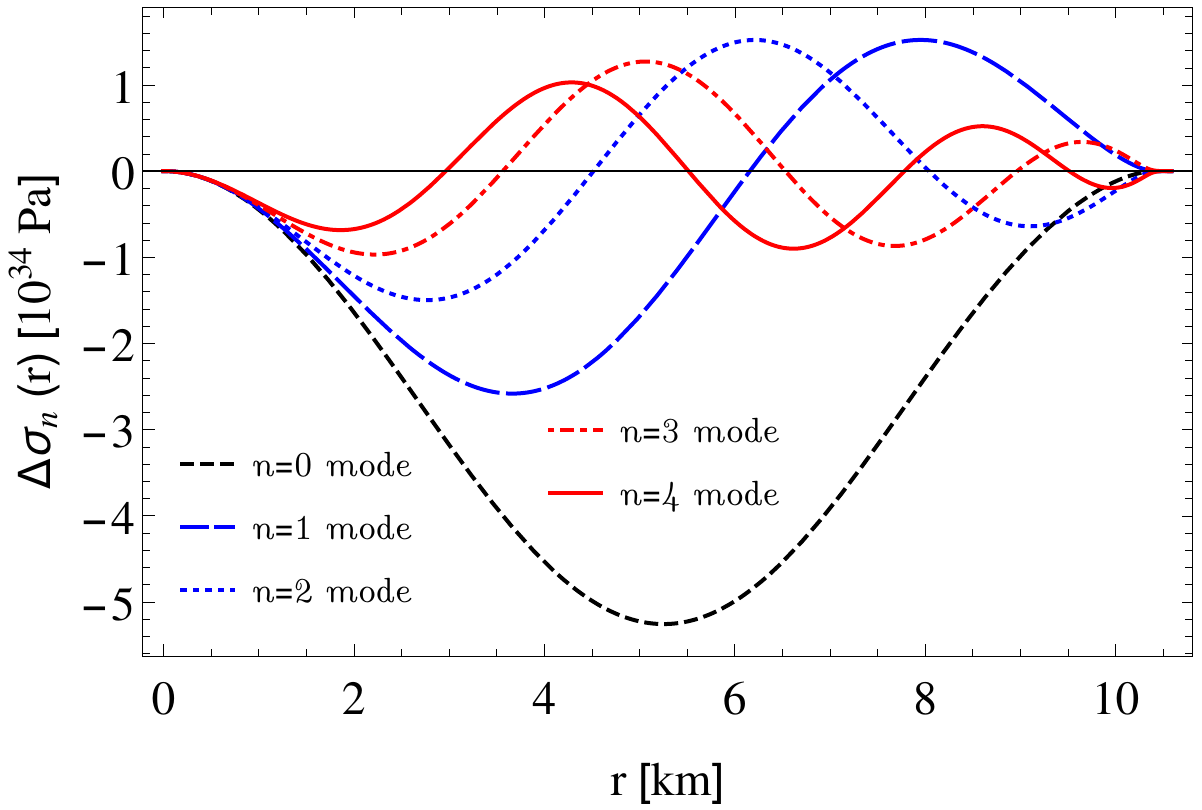} \
 \includegraphics[width=8.64cm]{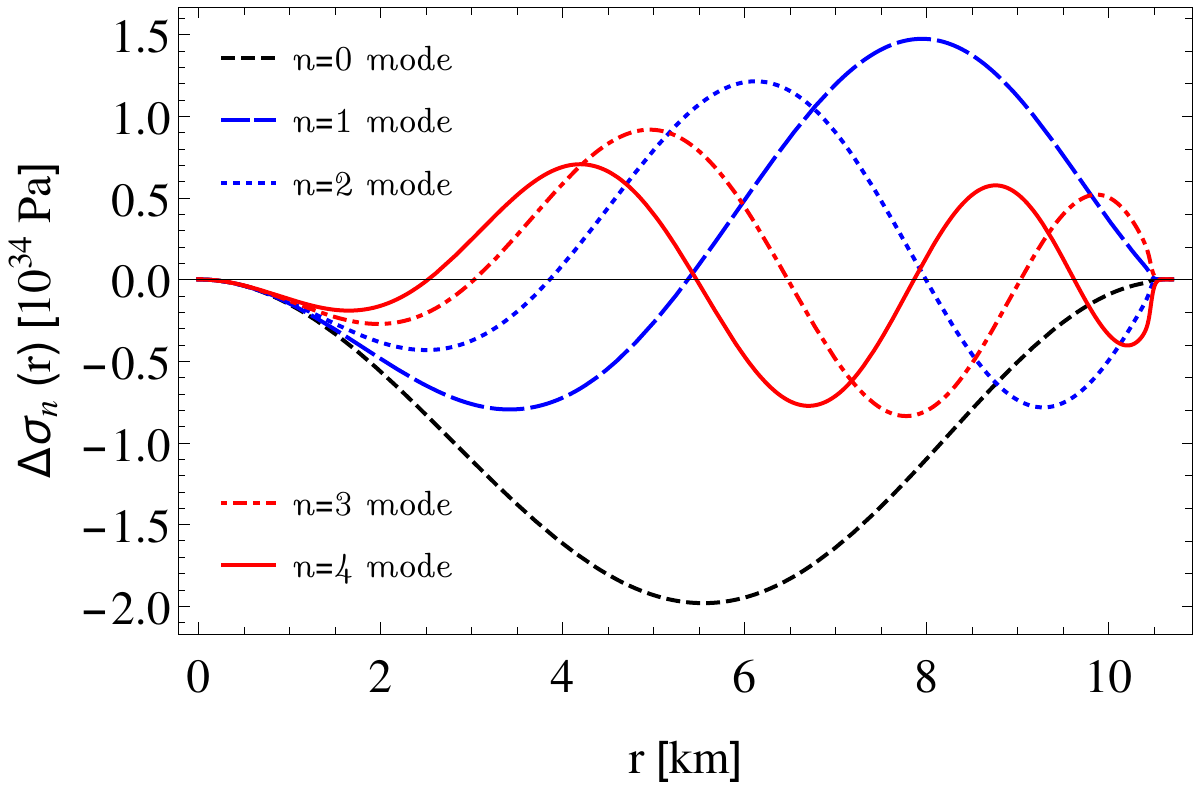}
 \includegraphics[width=8.4cm]{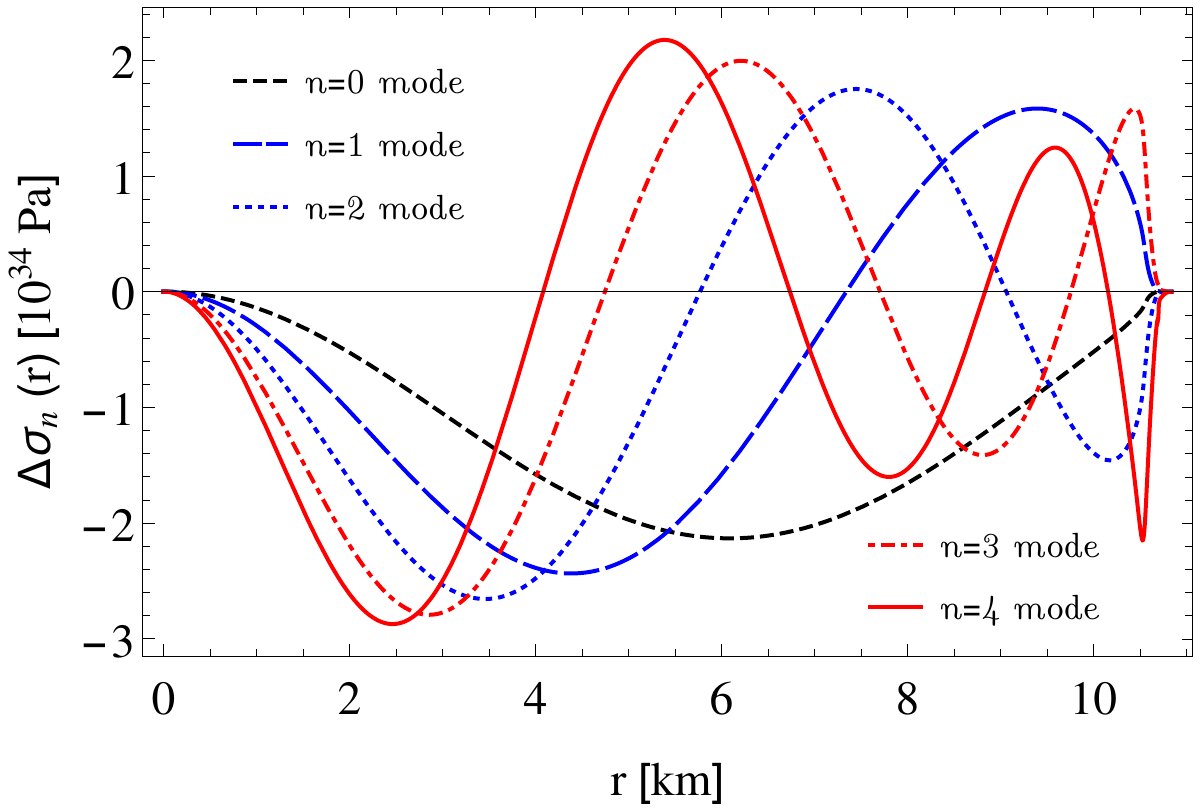} \
 \includegraphics[width=8.55cm]{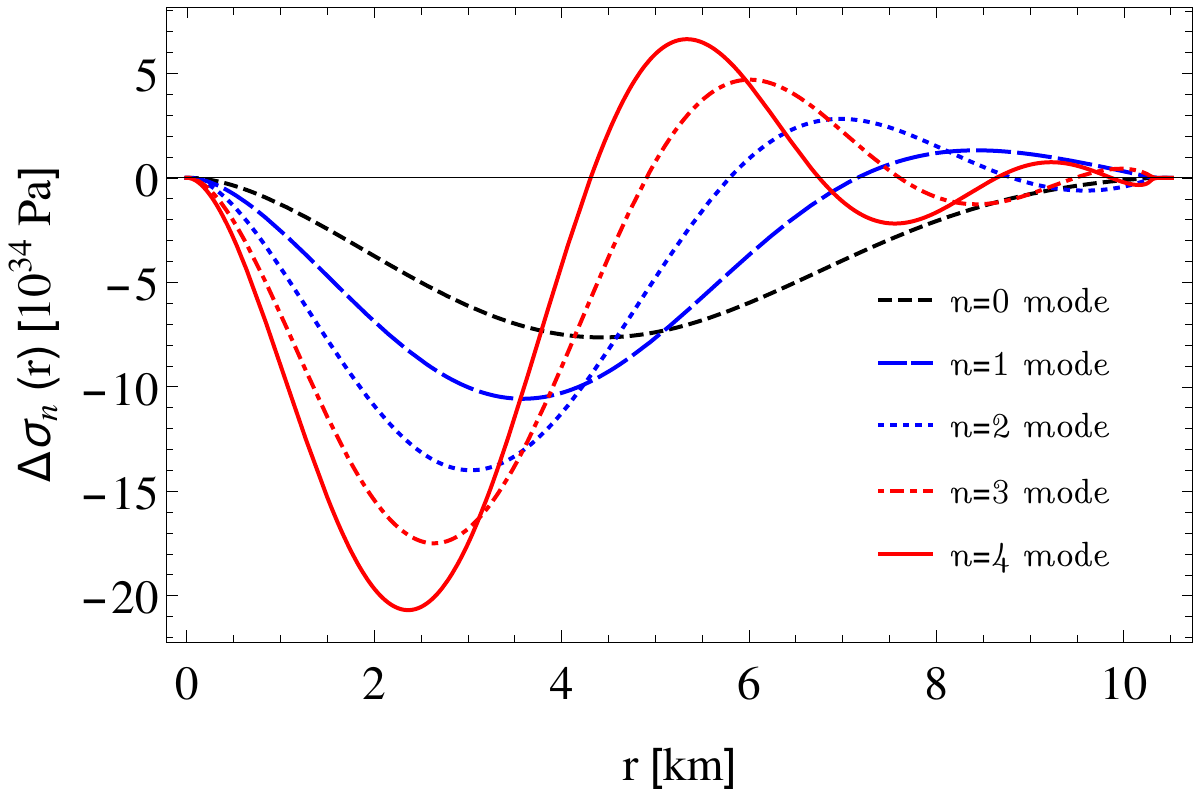}
 \caption{\label{figure3} Lagrangian perturbation of the anisotropy factor for the first five normal oscillation modes as a function of the radial coordinate obtained for a central mass density $\rho_c = 2.0 \times 10^{18}\ \rm kg/m^3$ with SLy EoS (\ref{29}). The upper left, upper right, lower left and lower right panels correspond to the anisotropy ansatze (\ref{30}) with $\beta_{\rm H} = 0.5$, (\ref{32}) with $\beta_{\rm BL} = 0.25$, (\ref{34}) with $\beta_{\rm HB} = -0.05$, and (\ref{36}) with $\beta_{\rm R} = -5.0\times 10^{11}\ \rm m^3$, respectively. The radius, mass, and frequency of the fundamental mode for such stars are shown in table \ref{table2}.}
\end{figure*}

Once the equilibrium quantities are known by solving the TOV equations, our second task is to verify if the $M(\rho_c)$ method is compatible with the calculation of frequencies. To that end, we have to solve the system of coupled first-order equations (\ref{21}) and (\ref{22}) with boundary conditions (\ref{23}) and (\ref{24}). The numerical solution of these equations is carried out using the shooting method, that is, we integrate the equations for a set of trial values of $\omega^2$ satisfying the condition (\ref{23}). Besides that, we consider that normalized eigenfunctions correspond to $\zeta(0) = 1$ at the center, and we integrate to the stellar surface. The values of the squared frequency for which the boundary condition (\ref{24}) is satisfied are the correct frequencies of the radial oscillations.  In particular, for a central mass density $\rho_c = 2.0 \times 10^{18}\ \rm{kg}/\rm{m}^3$ with anisotropy parameters $\beta_{\rm H} = 0.50$, $\beta_{\rm BL} = 0.25$, $\beta_{\rm HB} = -0.05$ and $\beta_{\rm R} = -5.0\times 10^{11}\ \rm m^3$, we display in Fig. \ref{figure2} the Lagrangian perturbation of the radial pressure for a set of test values $\omega^2$, where each minimum indicates the appropriate frequency. In other words, for a given stellar configuration there are different eigenvalues $\omega_n^2$ with their respective eigenfunctions $\zeta_n(r)$ and $\Delta p_{r,n}(r)$, where $n$ represents the number of nodes inside the anisotropic star. In Fig. \ref{figure2}, the first (leftmost) minimum represents the fundamental oscillation mode and it has no nodes between the center and the surface, whereas the first overtone $(n=1)$ has a node, the second overtone $(n=2)$ has two, and so forth. These radial pulsations in the anisotropy ansatz can be visualized in more detail in Fig. \ref{figure3}  and whose fundamental mode oscillation frequencies are shown in table \ref{table2}.

\begin{figure*} 
\includegraphics[width=8.41cm]{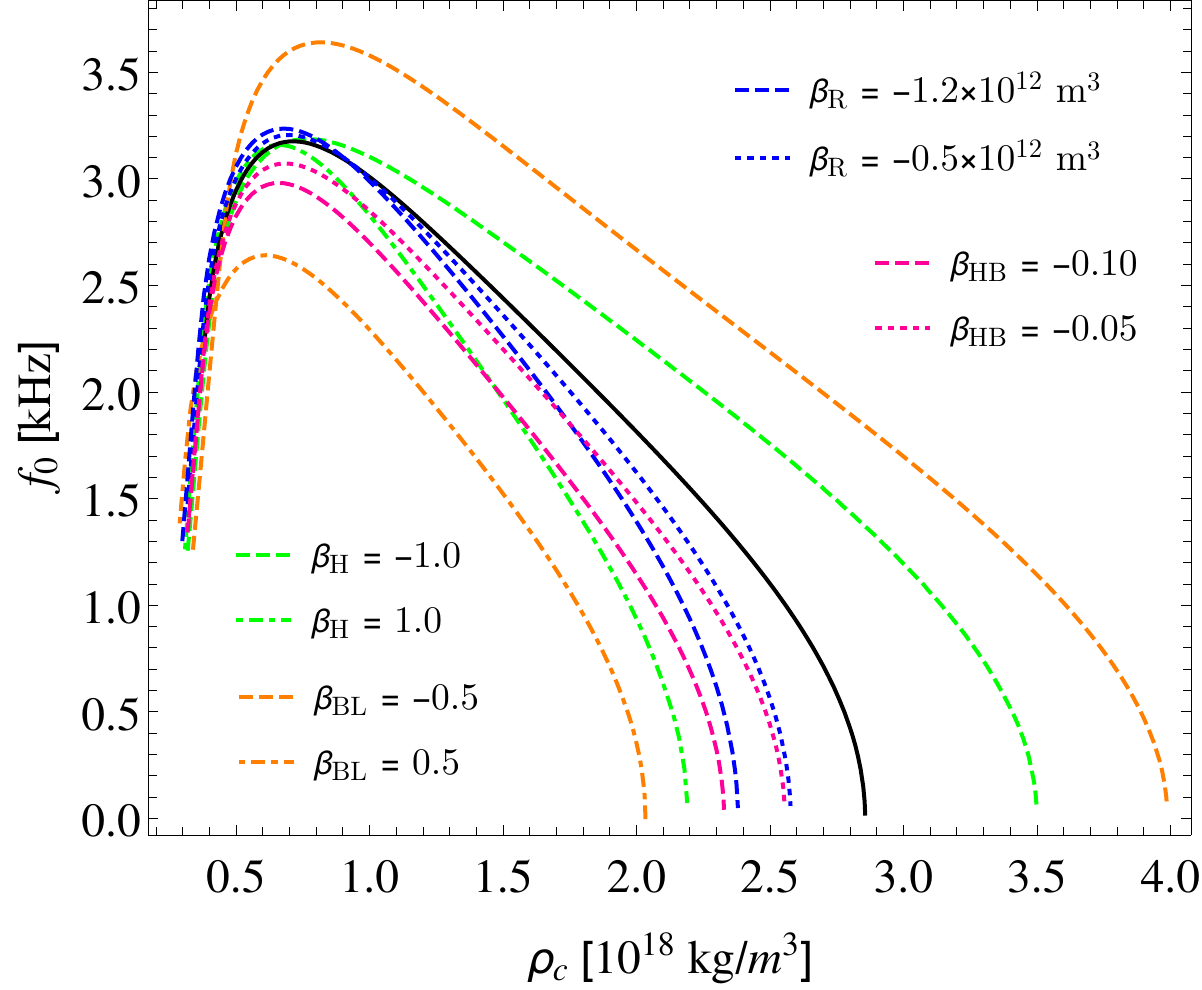} \
\includegraphics[width=8.58cm]{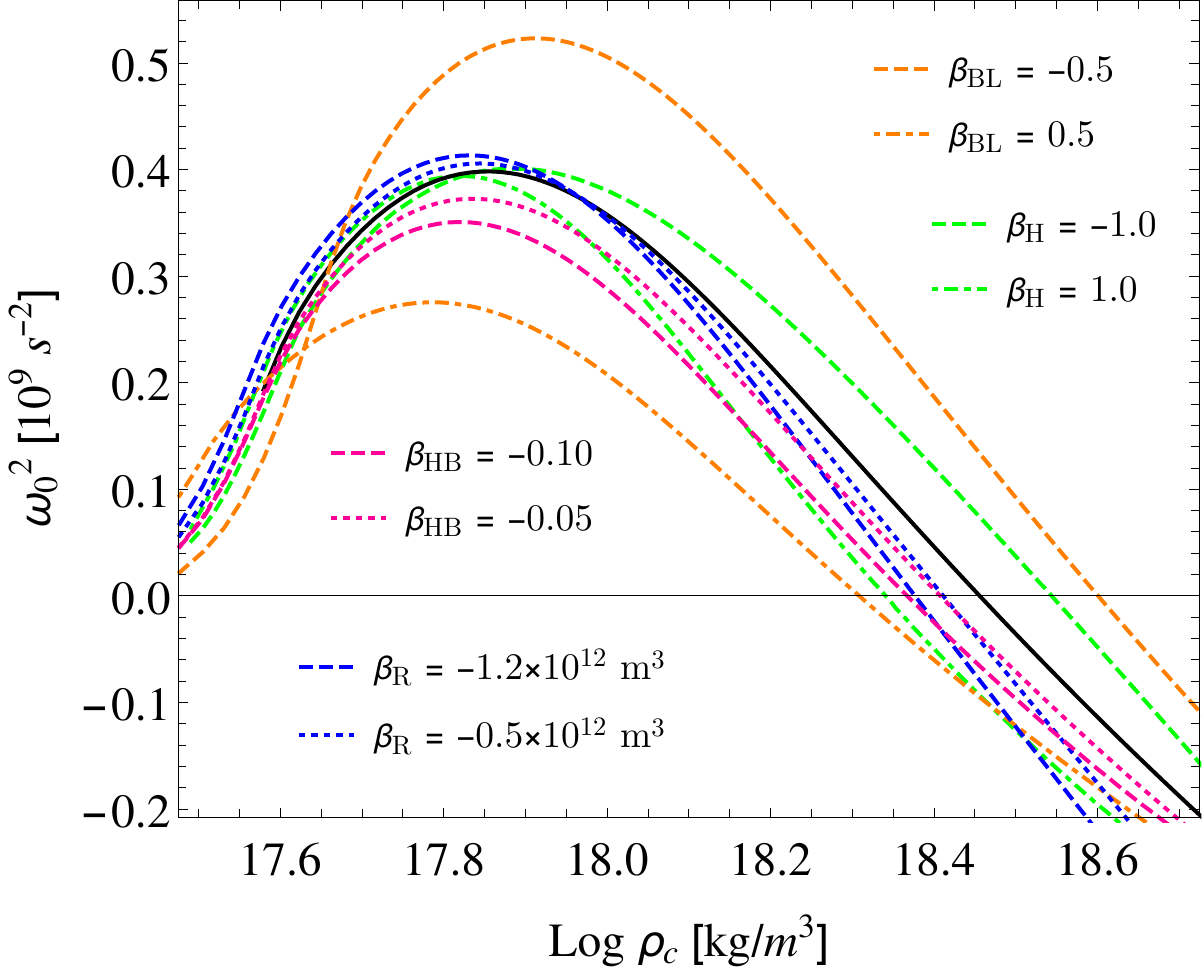} 
\caption{ On the left panel is shown the frequency of the fundamental mode $f_0 = \omega_0/2\pi$ as a function of the central mass density $\rho_c$, and on the right panel is displayed the squared frequency of the fundamental oscillation mode $\omega_0^2$ against the central density for different values of $\beta$. The isotropic case is shown in both plots as a benchmark by a solid black line.}
\label{figure4}
\end{figure*}

Using the central density as a parameter, the family of anisotropic stars that are actually stable is shown in the left plot of Fig. \ref{figure4}. The squared frequency of the fundamental oscillation mode against the central density is shown in the right plot of the same figure. Unlike the case of strange stars with MIT bag model EoS (where the frequency of the fundamental mode always decreases with increasing central density), in neutron stars $\omega_0^2$ increases to a maximum value and then decreases with $\rho_c$ regardless of the anisotropic model. In addition, for larger values of $\beta_{\rm H}$ and $\beta_{\rm BL}$, the onset of instability is indicated at a smaller and smaller central density value. On the other hand, for larger values of $\beta_{\rm HB}$ and $\beta_{\rm R}$, the onset of instability is found at a greater and greater central density value as we approach the isotropic case.

Taking into account the data recorded in table \ref{table1}, we see that only for the ansatze proposed by Horvat et al. (\ref{30}) and Raposo et al. (\ref{36}), the onset of instability indicated by the $M(\rho_c)$ method is located exactly at the configuration that has vanishing frequency of the fundamental mode. In other words, for these two anisotropy profiles, the maximum-mass point $M_{\rm max}$ and $\omega_0^2 = 0$ are reached at the same central density value. Nevertheless, for the anisotropic models suggested by Bowers-Liang (\ref{32}) and Herrera-Barreto (\ref{34}), the squared frequency of the fundamental mode does not pass through zero at the critical central density corresponding to the maximum-mass configuration. Therefore, it is evident that anisotropy affects the stellar stability and the critical central density does not always correspond to the onset of instability.

\begin{table*}
\caption{\label{table1}
Maximum-mass stellar configurations with SLy EoS for different values of the anisotropy parameter. The mass density values correspond to the critical central density where the function $M(\rho_c)$ is a maximum. The fundamental mode frequency with an asterisk means that it is imaginary, and $z(R) = e^{\lambda(R)} -1$ is the gravitational redshift of light emitted at the surface of the equilibrium star.}
\begin{ruledtabular}
\begin{tabular}{lccccc}
Free parameter  &  $\rho_{c}$ [$10^{18}\ {\rm kg}/ {\rm m}^3$]  &  $R$  [\rm{km}]  &  $M$ [$M_\odot$]   &   $z(R)$   &  $f_0$ [kHz]  \\
\colrule
Isotropic case  & 2.857  &  9.982  &  2.046  &  0.593  &  0.00  \\
$\beta_{\rm H} = -1.00$  &  3.501  &  9.784  &  1.762  &  0.462  &  0.00  \\
$\beta_{\rm H} = -0.50$  &  3.199  &  9.870  &  1.903  &  0.525  &  0.00  \\
$\beta_{\rm H} = 0.50$  &  2.512  &  10.122  &  2.190  &  0.665 &  0.00  \\
$\beta_{\rm H} = 1.00$  &  2.191  &  10.292  &  2.330  &  0.739  &  0.00 \\
$\beta_{\rm BL} = -0.50$  &  3.457  &  9.614  &  1.837  &  0.516 &  1.197  \\
$\beta_{\rm BL} = -0.25$  &  3.149  &  9.787  &  1.936  &  0.552  &  0.811  \\
$\beta_{\rm BL} = 0.25$  &  2.579  &  10.203  &  2.171  &  0.641 &  0.730*  \\
$\beta_{\rm BL} = 0.50$  &  2.313  &  10.460  &  2.312  &  0.699  &  0.969*  \\
$\beta_{\rm HB} = -0.10$  &  2.459  &  10.544  &  2.308  &  0.683  &  0.674*  \\
$\beta_{\rm HB} = -0.05$  &  2.641  &  10.273  &  2.182  &  0.639  &  0.531*  \\
$\beta_{\rm R} = -1.0\times 10^{12}\ \rm m^3$  &  2.429  &  9.955  &  2.262  &  0.745  &  0.00  \\
$\beta_{\rm R} = -0.5\times 10^{12}\ \rm m^3$  &  2.578  &  9.930  &  2.169  &  0.680  &  0.00  \\
\end{tabular}
\end{ruledtabular}
\end{table*}

\begin{table}
\caption{\label{table2} 
Anisotropic neutron stars with central mass density $\rho_c = 2.0 \times 10^{18}\ \rm{kg}/\rm{m}^3$  for some anisotropy parameters. The first five eigenfunctions for oscillation modes of these configurations are shown in Fig. \ref{figure3}.}
\begin{ruledtabular}
\begin{tabular}{lccc}
Free parameter  &  $R$  [\rm{km}]  &  $M$ [$M_\odot$]  &  $f_0$ [kHz]  \\
\colrule
$\beta_{\rm H} = 0.50$  &  10.630  &  2.161  &  1.467  \\
$\beta_{\rm BL} = 0.25$  &  10.745  &  2.138  &  1.264  \\
$\beta_{\rm HB} = -0.05$  &  10.860  &  2.144  &  1.482  \\
$\beta_{\rm R} = -5.0\times 10^{11}\ \rm m^3$  &  10.547  &  2.131  &  1.623  \\
\end{tabular}
\end{ruledtabular}
\end{table}

\subsection{Dynamical evolution of gravitational collapse}

Anisotropic neutron stars in hydrostatically stable equilibrium oscillate with a purely real (fundamental) frequency when are subjected to a radial perturbation, whereas the unstable stars (with imaginary frequency of the lowest oscillation mode) undergo a gravitational collapse from rest to form a black hole. We assume that the unstable configurations are initially in a state of hydrostatic equilibrium and then gradually begin to collapse until the formation of an event horizon. In the case of unstable anisotropic neutron stars with SLy EoS and anisotropic model proposed by Bowers and Liang (\ref{32}), for an initial central mass density $\rho_c = 2.6 \times 10^{18}\ \rm{kg}/\rm{m}^3$ and anisotropy parameter $\beta_{\rm BL} = 0.2$, we solve Eq. (\ref{57}) with final conditions $f = 0.150$ and $df/dt = -8.613 \times 10^3\ \rm{s}^{-1}$ at time $t =t_{bh} = -1.723 \times 10^{-5}\ \rm{s}$. For this particular configuration, the tangential pressure dominates the radial pressure and $\omega_0^2 = -13.001\times 10^6 \ \rm s^{-2}$. Then we perform a time displacement so that this unstable star evolves from the initial instant $t = 0$ (when the interior structure is governed by the background equations and the external solution is Schwarzschild-type) until the moment of horizon formation $t_{bh} = 1.981\ \rm{ms}$, that is, when the star has collapsed and the mass of the resulting black hole is $m_{bh} = 1.335\ M_\odot$.

\begin{figure} 
\includegraphics[width=8.2cm]{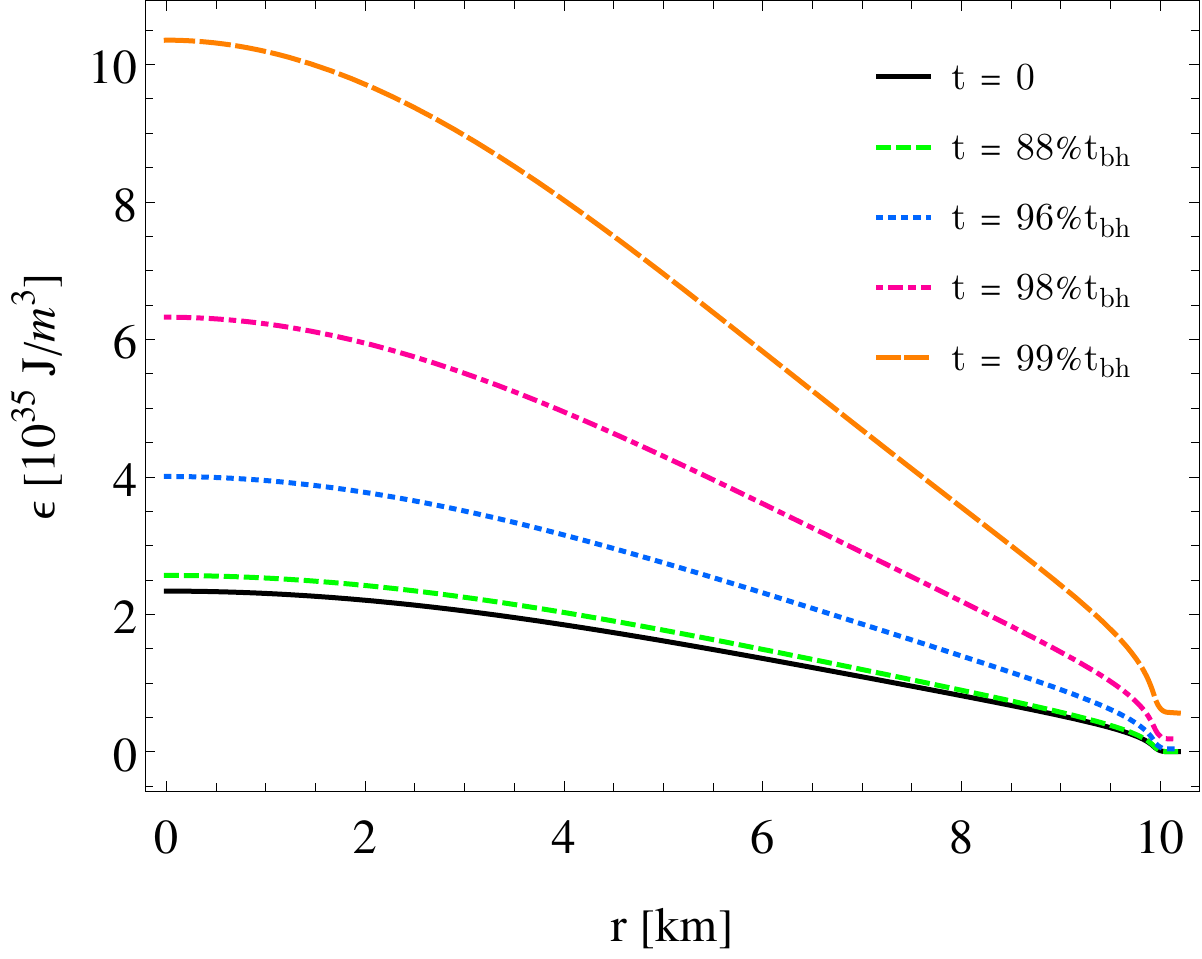} 
\includegraphics[width=8.2cm]{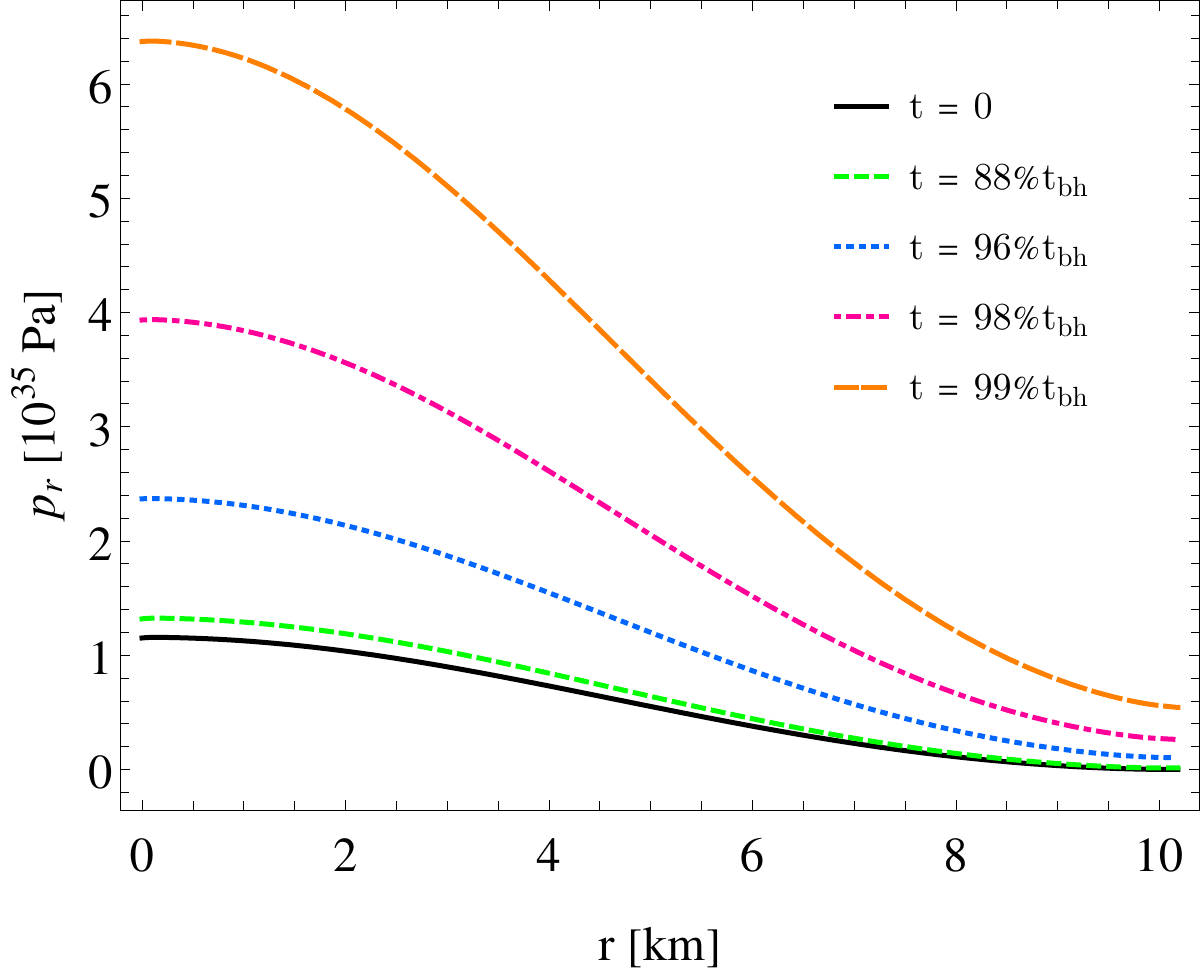} 
\caption{ Energy density (upper panel) and radial pressure (lower panel) as functions of the radial coordinate at different times, for a central mass density $\rho_c = 2.6 \times 10^{18}\ \rm{kg}/\rm{m}^3$ with anisotropy ansatz (\ref{32}) for $\beta_{\rm BL} = 0.2$. This configuration corresponds to an unstable anisotropic neutron star with radius $R = 10.183\ \rm{km}$, initial mass $M = 2.145\ M_\odot$ and $m_{bh} = 1.335\ M_\odot$ at the end of the collapse.}
\label{figure5}
\end{figure}

\begin{figure} 
\includegraphics[width=8.2cm]{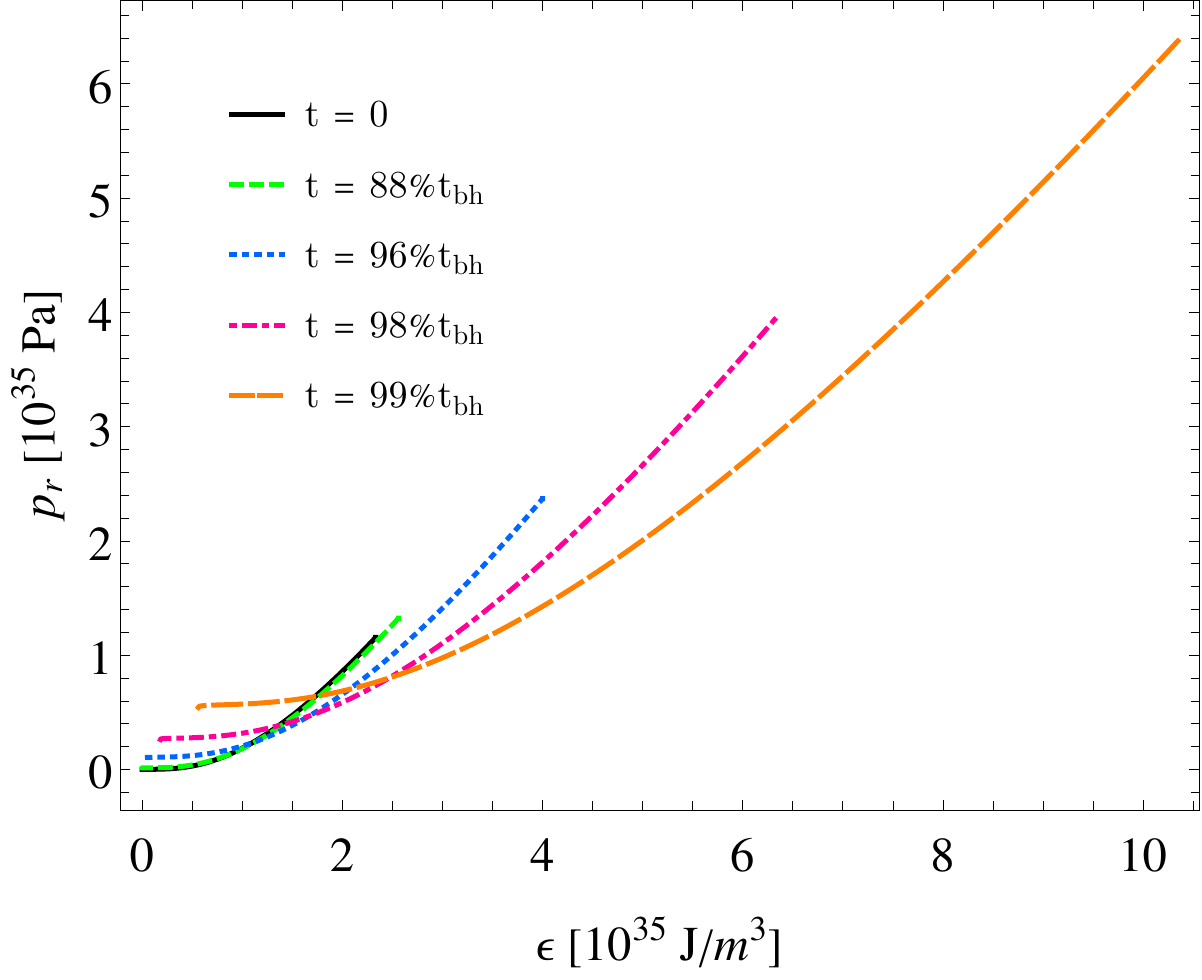} 
\includegraphics[width=8.2cm]{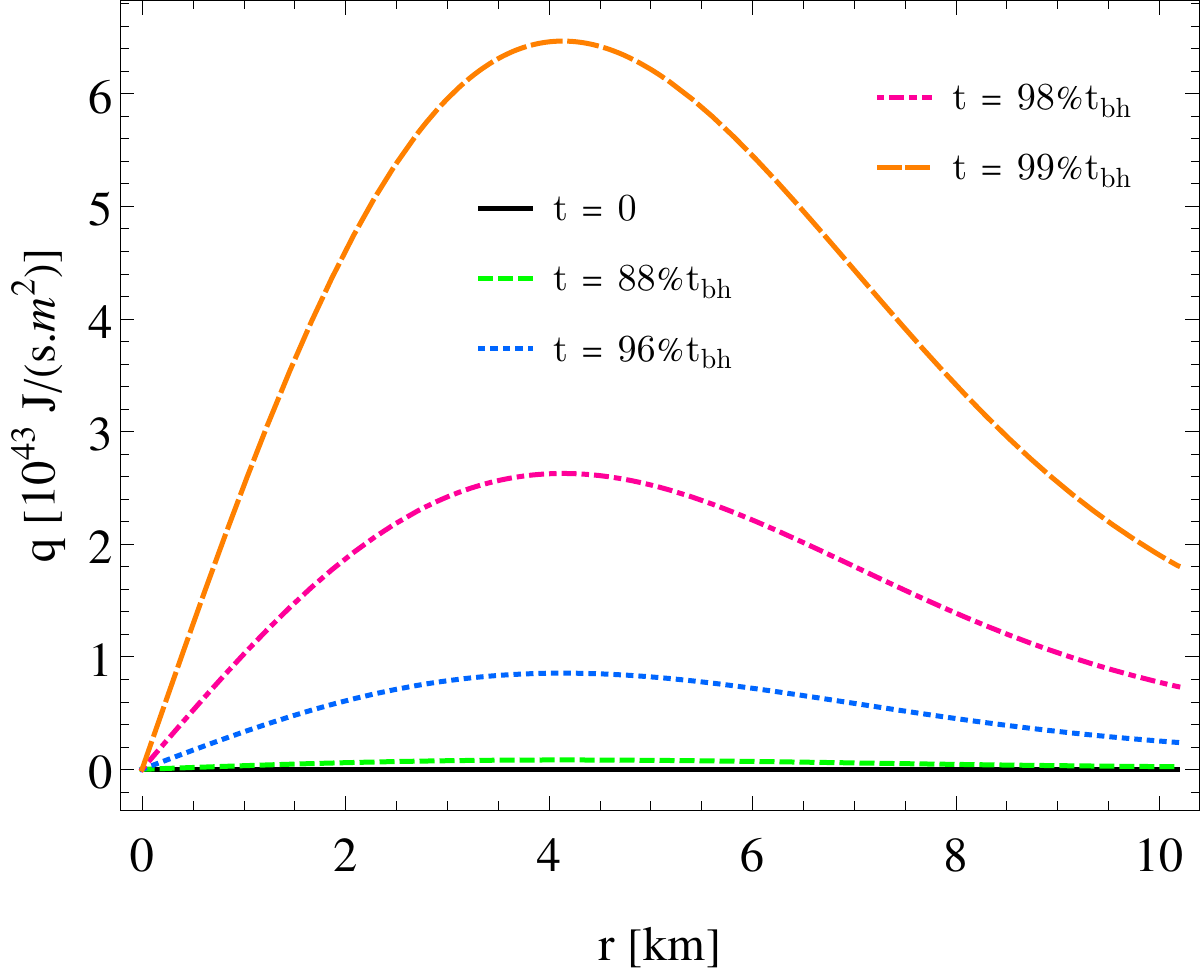} 
\caption{ Upper panel: Evolution of the EoS $p_r = p_r(\epsilon)$ during the gravitational collapse. Lower panel: Heat flux as a function of the radial coordinate at different times. The central density and anisotropy parameter have the same values as in Fig. \ref{figure5}.}
\label{figure6}
\end{figure}

The energy density (\ref{62}) and radial pressure (\ref{63}) as functions of the radial coordinate at different times are displayed in the upper and lower panels of Fig. \ref{figure5}, respectively. Both thermodynamic quantities present their maximum values at the stellar center and change significantly in the last moments of the collapse, while near the surface the changes are relatively small. Unlike the static case, the radial pressure at the surface is no longer zero during the dynamical evolution of the gravitational collapse  because there exist a radial heat flux according to Eq. (\ref{56}). As a result, we can investigate how the EoS behaves as the star collapses. The upper panel of Fig. \ref{figure6} reveals that the EoS for the radial pressure undergoes sudden changes during the collapse of an unstable neutron star, where the maximum and minimum values in each curve correspond to the center and the surface of the star, respectively.

\begin{figure} 
\includegraphics[width=8.2cm]{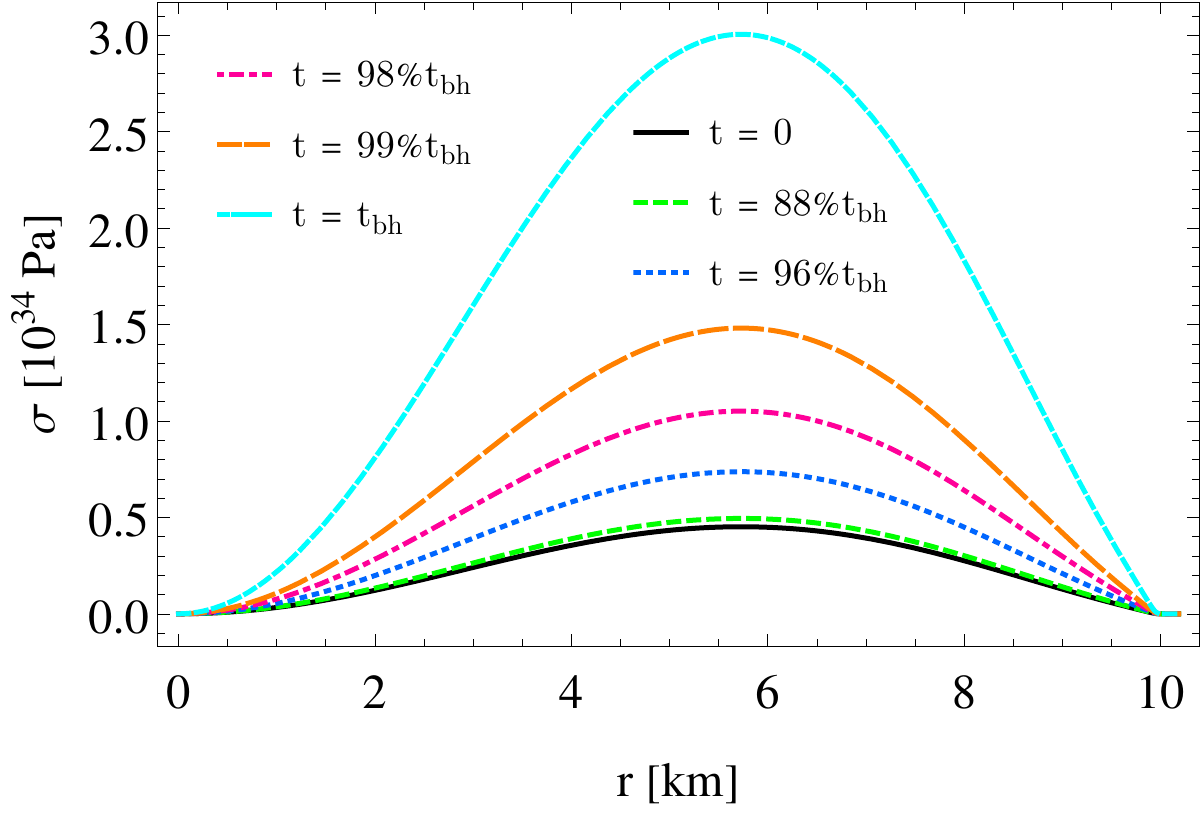} 
\includegraphics[width=8.2cm]{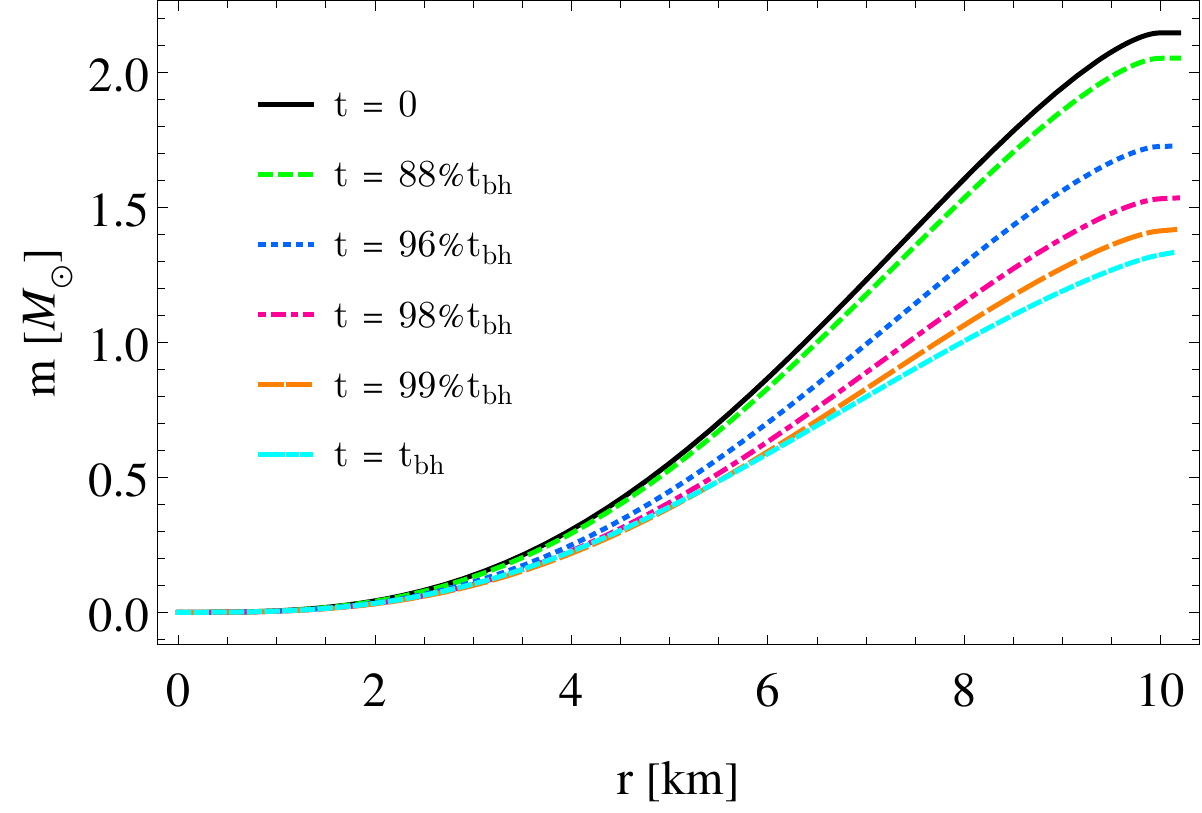} 
\includegraphics[width=8.2cm]{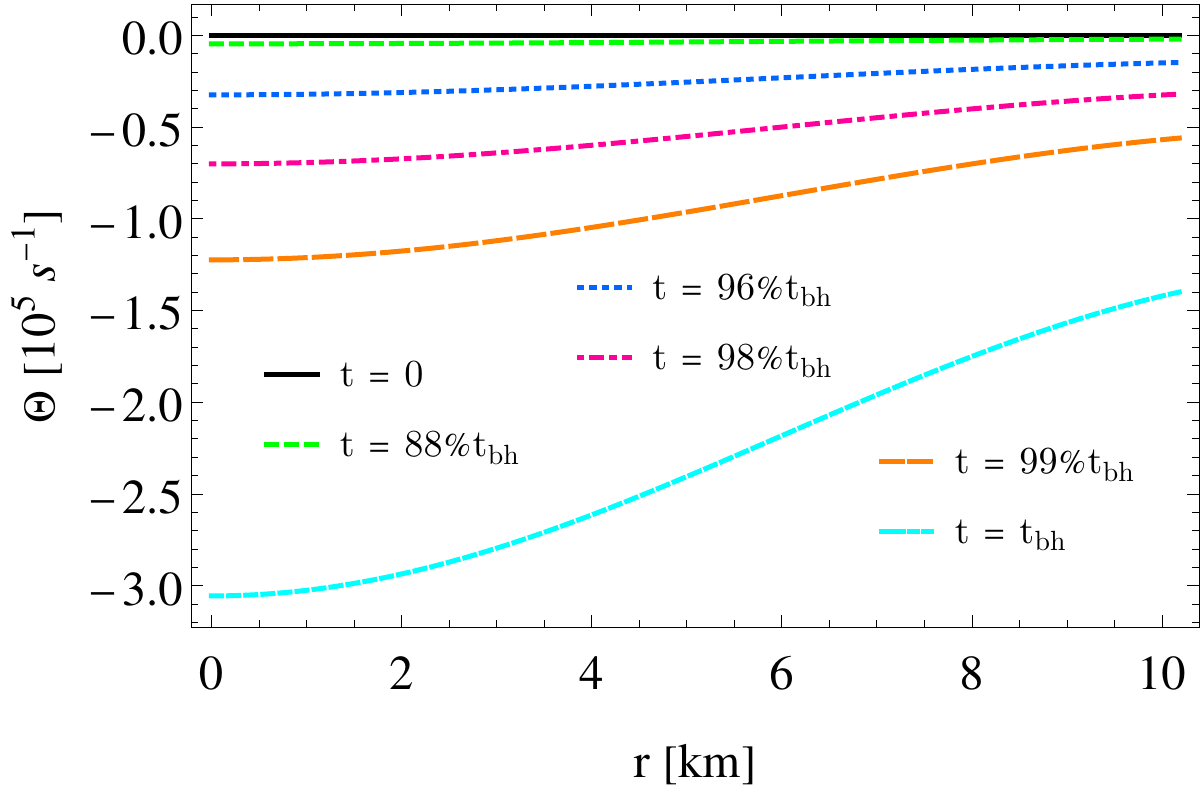} 
\caption{ Radial profile of the anisotropy ansatz (upper panel) described by Eq. (\ref{66}), mass function (intermediate panel) given by (\ref{46}), and expansion scalar (lower panel) during the collapse process at different instants of time. The results are for the same solution as in Fig. \ref{figure5}.}
\label{figure7}
\end{figure}

In the lower panel of Fig. \ref{figure6} we plot the radial heat flux (\ref{65}), which undergoes great alterations in the intermediate regions of the collapsing configuration and its value is not zero at the surface. Indeed, when the heat flux vanishes, the radial pressure also vanishes at the surface and the exterior solution is the Schwarzschild vacuum solution. According to the upper panel of Fig. \ref{figure7}, the degree of anisotropy in the pressures increases as the star collapses and it always vanishes at the origin as well as at the surface for any instant of time. The radial profile of the mass function (\ref{46}) is displayed in the intermediate panel of the same figure, indicating that it decreases during the gravitational collapse due to the emission of particles into outer spacetime. On the stellar surface and at the moment of event horizon formation, we have $m(t_{bh}, R) = 1.335\ M_\odot$ which precisely coincides with the value obtained by means of the junction condition (\ref{61}). The masses corresponding to the black hole formed by the gravitational collapse of anisotropic neutron stars for some central density values are shown in tables \ref{table3} and \ref{table4} of Appendix \ref{ApendixB}.

In the lower panel of Fig. \ref{figure7} we illustrate the radial behaviour of the expansion scalar (\ref{67}) during the collapse process. For any instant of time, it can be seen that $\Theta < 0$ and $\partial\Theta/\partial r > 0$, which means that the stellar system is collapsing. In addition, it is clear that $\Theta \rightarrow 0$ as $t \rightarrow 0$.

At long last, the collapsing anisotropic neutron star with initial central mass density $\rho_c = 2.6 \times 10^{18}\ \rm{kg}/\rm{m}^3$, SLy EoS for radial pressure and anisotropy parameter $\beta_{\rm BL} = 0.2$, is physically reasonable because it obeys the energy conditions in the full extent of the star and throughout the collapse process. It is worth emphasizing that we have tested this procedure for the other anisotropy ansatze (\ref{30}), (\ref{34}) and (\ref{36}), obtaining a similar behaviour during the dynamical evolution of the collapse.

%--------------------------------------------------------------

\section{Conclusions}\label{Sec5}

In this paper, we have constructed families of anisotropic neutron stars for an EoS compatible with the recent observation of the event GW170817 and for four different anisotropy ansatze, aiming to settle bounds on the maximum masses reachable in this kind of anisotropic models. We have carried out an analysis of adiabatic radial pulsations for such stars in order to study their radial stability against gravitational collapse. We have also developed a dynamical model that describes the non-adiabatic gravitational collapse of the unstable anisotropic configurations, and TOV equations have been generalized within this context. To summarize, the main conclusions of this work are:

\begin{itemize}
\item[•] For the SLy EoS in the radial pressure, we have constrained the anisotropy parameters provided in the literature in order to satisfy the recent restriction of the maximum mass of neutron stars based on gravitational wave observations. Our results thus suggest that  $\beta_{\rm H} \lesssim 0.40$, $\beta_{\rm BL} \lesssim 0.23$, $\beta_{\rm HB} \gtrsim -0.042$, and $\beta_{\rm R} \gtrsim -4.6 \times 10^{11}\ \rm m^3$. However, we must point out that these bounds can be slightly altered for other EoSs.

\item[•] Anisotropy affects the stellar stability and the critical central density (where $dM/d\rho_c =0$) does not always correspond to the onset of instability. In particular, the maximum point on the $M(\rho_c)$ curve does not indicate the onset of instability for the anisotropic models proposed by Bowers-Liang \cite{BowersLiang} and Herrera-Barreto \cite{HerreraBarreto}. Nevertheless, we remark that this criterion is compatible with the calculation of the oscillation frequencies in the case of the ansatze suggested by Horvat et al. \cite{Horvat2011} and Raposo et al. \cite{Raposo2019}.

\item[•] Given an initial value of central mass density and a specific anisotropy profile for an unstable anisotropic neutron star, we have investigated the evolution of the equation of state for radial pressure and anisotropy ansatz as the star undergoes a non-adiabatic gravitational collapse. The sudden changes in all relevant physical quantities occur near the formation of the event horizon as a consequence of a radial heat flow.
\end{itemize}

\begin{acknowledgments}
JMZP acknowledges Brazilian funding agency CAPES for PhD scholarship 331080/2019.
\end{acknowledgments}\

\appendix

\section{Extrinsic curvature to $\Sigma$}\label{ApendixA}

Here we present the non-vanishing components of the extrinsic curvature tensor (\ref{52}) used in Sec. \ref{Sec3}:
\begin{eqnarray}
K_{\tau\tau}^- &=& -\left[ \left( \frac{dt}{d\tau} \right)^2 \frac{d\psi}{dr} \frac{e^{2\psi -\lambda}}{\sqrt{f}} \right]_\Sigma ,  \\
K_{\theta\theta}^- &=& \left[ \frac{r\sqrt{f}}{e^\lambda} \right]_\Sigma ,  \\
K_{\phi\phi}^- &=& K_{\theta\theta}^- \sin^2\theta ,  \\
K_{\tau\tau}^+ &=& \left[ \frac{1}{c}\frac{d^2\upsilon}{d\tau^2}\left( \frac{d\upsilon}{d\tau} \right)^{-1} - \frac{Gm}{c^2\chi^2}\left( \frac{d\upsilon}{d\tau} \right) \right]_\Sigma ,  \qquad  \\
K_{\theta\theta}^+ &=& \left[ \frac{d\upsilon}{d\tau}\left( 1 -\frac{2Gm}{c^2\chi} \right)\chi + \frac{\chi}{c}\frac{d\chi}{d\tau} \right]_\Sigma ,  \\
K_{\phi\phi}^+ &=& K_{\theta\theta}^+ \sin^2\theta .
\end{eqnarray}

\section{Parameters and oscillation spectrum of anisotropic neutron stars}\label{ApendixB}

Because of the $M(\rho_c)$ method is not compatible with the calculation of the oscillation frequencies for the anisotropic models proposed by Bowers-Liang and Herrera-Barreto, in this appendix we provide two tables of numerical data corresponding to anisotropy ansatze (\ref{32}) and (\ref{34}) with SLy EoS for radial pressure. For some values of central mass density, we present the radius, total mass, frequency of the fundamental mode and the first overtone, as well as the mass of the formed black hole for the unstable anisotropic configurations. In the case of unstable stars, when $\omega_0^2 <0$, the frequency is imaginary and we are denoting it by an asterisk.

\begin{table}[t]
\caption{\label{table3} 
Data for $\beta_{\rm BL} = 0.2$ with SLy EoS in the radial pressure and anisotropy ansatz (\ref{32}).}
\begin{ruledtabular}
\begin{tabular}{cccccc}
$\rho_{c}$    &    $R$    &    $M$   &   $f_0$   &    $f_1$   &   $m_{bh}$  \\
$[10^{18} \rm{kg}/ \rm{m}^3]$   &   [\rm{km}]   &   [$M_\odot$]   &   [kHz]   &   [kHz]   &   [$M_\odot$]   \\
\colrule
  0.800  &  12.018  &  1.228  &  2.927  &  6.525  &  --  \\
  1.000  &  11.872  &  1.543  &  2.744  &  6.718  &  --  \\
  1.500  &  11.299  &  1.965  &  2.106  &  6.584  &  --  \\
  2.000  &  10.731  &  2.108  &  1.386  &  6.302  &  --  \\
  2.400  &  10.349  &  2.141  &  0.588  &  6.090  &  --  \\
  2.600  &  10.183  &  2.145  &  0.574*  &  5.992  &  1.335  \\
  3.000  &  9.892  &  2.138  &  1.228*  &  5.813  &  1.366  \\
  3.200  &  9.765  &  2.131  &  1.423*  &  5.732  &  1.374  \\
  3.400  &  9.648  &  2.122  &  1.583*  &  5.655  &  1.379  \\
  3.600  &  9.541  &  2.112  &  1.718*  &  5.582  &  1.382  \\
\end{tabular}
\end{ruledtabular}
\end{table}

\begin{table}
\centering
\caption{\label{table4} 
Data for $\beta_{\rm BH} = -0.04$ with SLy EoS in the radial pressure and anisotropy ansatz (\ref{34}).}
\begin{ruledtabular}
\begin{tabular}{cccccc}
$\rho_{c}$    &    $R$    &    $M$   &   $f_0$   &    $f_1$   &   $m_{bh}$  \\
$[10^{18} \rm{kg}/ \rm{m}^3]$   &   [\rm{km}]   &   [$M_\odot$]   &   [kHz]   &   [kHz]   &   [$M_\odot$]   \\
\colrule
  0.800  &  12.147  &  1.224  &  3.056  &  6.607  &  --  \\
  1.000  &  11.979  &  1.540  &  2.880  &  6.862  &  --  \\
  1.500  &  11.393  &  1.967  &  2.248  &  6.751  &  --  \\
  2.000  &  10.824  &  2.115  &  1.549  &  6.464  &  --  \\
  2.400  &  10.445  &  2.151  &  0.861  &  6.245  &  --  \\
  2.600  &  10.279  &  2.155  &  0.182  &  6.143  &  --  \\
  3.000  &  9.991  &  2.151  &  1.095*  &  5.958  &  1.369  \\
  3.200  &  9.865  &  2.145  &  1.319*  &  5.873  &  1.378  \\
  3.400  &  9.749  &  2.137  &  1.496*  &  5.794  &  1.384  \\
  3.600  &  9.643  &  2.127  &  1.643*  &  5.718  &  1.387  \\
\end{tabular}
\end{ruledtabular}
\end{table}


\newpage

\end{document}